\documentclass[sigconf]{acmart}

\usepackage{multirow}
\usepackage{enumitem}
\usepackage{color, colortbl}

\AtBeginDocument{%
  \providecommand\BibTeX{{%
    \normalfont B\kern-0.5em{\scshape i\kern-0.25em b}\kern-0.8em\TeX}}}

\newcommand{\wh}{Wheeler}
\newcommand{\sysname}{Wheeler}
\newcommand{\wa}{\texttt{\small Wheel-1}}
\newcommand{\wb}{\texttt{\small Wheel-2}}
\newcommand{\wc}{\texttt{\small Wheel-3}}
\newcommand{\hnav}{\texttt{\small H-nav}}
\newcommand{\fnav}{\texttt{\small 2d-nav}} 
\newcommand{\tnav}{\texttt{\small 2d-T-nav}} 
\newcommand{\parti}{participant}
\newcommand{\partis}{participants}

\newcommand{\Partis}{Participants}

\def \billahdebug{}

\ifx \billahdebug \undefined
\newcommand{\fixmesb}[1]{{}}
\newcommand{\fixmeth}[1]{{}}
\newcommand{\fixmens}[1]{{}}
\newcommand{\fixmera}[1]{{}}
\newcommand{\new}[1]{{}}

\else

\newcommand{\T}[1]{{\texttt{\small #1}}}

\newcommand{\fx}[1]{{\bf\textcolor{red}{ [ #1 ]}}}
\newcommand{\up}[1]{{\bf\textcolor{blue}{~(\faThumbsUp)~}}}
\newcommand{\down}[1]{{\bf\textcolor{red}{~(\faThumbsDown)~}}}

\newcommand{\fixmesb}[1]{{\bf\textcolor{red}{ [ sB FIXME: #1 ]}}}
\newcommand{\fixmeth}[1]{{\bf\textcolor{blue}{ [ tH FIXME: #1 ]}}}

\newcommand{\new}[1]{{\textcolor{black}{#1}}}
\fi
\newcolumntype{L}[1]{>{\raggedright\let\newline\\\arraybackslash\hspace{0pt}}m{#1}}
\newcolumntype{C}[1]{>{\centering\let\newline\\\arraybackslash\hspace{0pt}}m{#1}}
\newcolumntype{R}[1]{>{\raggedleft\let\newline\\\arraybackslash\hspace{0pt}}m{#1}}
\copyrightyear{2024}
\acmYear{2024}
\setcopyright{acmlicensed}\acmConference[UIST '24]{The 37th Annual ACM Symposium on User Interface Software and Technology}{October 13--16, 2024}{Pittsburgh, PA, USA}
\acmBooktitle{The 37th Annual ACM Symposium on User Interface Software and Technology (UIST '24), October 13--16, 2024, Pittsburgh, PA, USA}
\acmDOI{10.1145/3654777.3676396}
\acmISBN{979-8-4007-0628-8/24/10}

\begin{document}

\title[Wheeler: A Three-Wheeled Input Device]{Wheeler: A Three-Wheeled Input Device for Usable, Efficient, and Versatile Non-Visual Interaction}

\author[MT Islam]{Md Touhidul Islam$^*$}
\affiliation{%
  \institution{Pennsylvania State University}  
  \city{University Park}
  \state{PA}
  \country{USA}
}
\email{touhid@psu.edu}

\author[N Sojib]{Noushad Sojib$^*$}
\affiliation{%
  \institution{University of New Hampshire}  
  \city{Durham}
  \state{NH}
  \country{USA}
}
\email{noushad.sojib@unh.edu}

\author[I Kabir]{Imran Kabir}
\affiliation{%
  \institution{Pennsylvania State University}  
  \city{University Park}
  \state{PA}
  \country{USA}
}
\email{ibk5106@psu.edu}

\author[AR Amit]{Ashiqur Rahman Amit}
\affiliation{%
  \institution{Innovation Garage Limited}  
  \city{Dhaka}
  \country{Bangladesh}
}
\email{amit@innovationgarage.com.bd}

\author[M Ruhul Amin]{Mohammad Ruhul Amin}
\affiliation{%
  \institution{Fordham University}  
  \city{Bronx}
  \state{NY}
  \country{USA}
}
\email{mamin17@fordham.edu}

\author[SM Billah]{Syed Masum Billah}
\affiliation{%
  \institution{Pennsylvania State University}  
  \city{University Park}
  \state{PA}
  \country{United States}
}
\email{sbillah@psu.edu}







\thanks{$^*$Equal Contribution}

\begin{abstract}
Blind users rely on keyboards and assistive technologies like screen readers to interact with user interface (UI) elements. In modern applications with complex UI hierarchies, navigating to different UI elements poses a significant accessibility challenge. Users must listen to screen reader audio descriptions and press relevant keyboard keys one at a time. This paper introduces Wheeler, a novel three-wheeled, mouse-shaped stationary input device, to address this issue. Informed by participatory sessions, Wheeler enables blind users to navigate up to three hierarchical levels in an app independently using three wheels instead of navigating just one level at a time using a keyboard. The three wheels also offer versatility, allowing users to repurpose them for other tasks, such as 2D cursor manipulation. A study with 12 blind users indicates a significant reduction (40\%) in navigation time compared to using a keyboard. Further, a diary study with our blind co-author highlights Wheeler's additional benefits, such as accessing UI elements with partial metadata and facilitating mixed-ability collaboration.
\end{abstract}



\begin{teaserfigure}
    \centering
    \includegraphics[width=.99\textwidth]{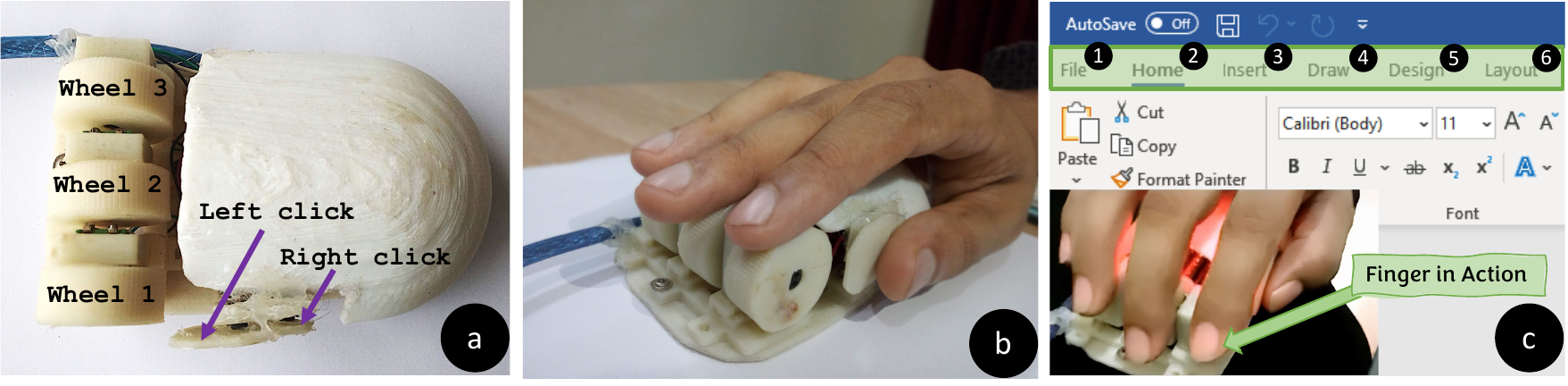}
    \caption{
     \textbf{{\sc \textbf{\sysname{}}} input device and its usage in the wild}.  (a) shows a 3D-printed implementation of \wh{} having three wheels and two push buttons (primary and secondary) on the side;
    (b) shows a blind user holding the device, placing three central fingers on the three wheels, and the thumb over the two side buttons;
    (c) shows how a blind user can navigate the multi-level, hierarchical menu (e.g., Microsoft Word's ribbon menu) using \sysname{}'s \hnav{} mode.
    The pale green rectangle at the top shows the first level menu in the app, with the numbers 1-6 each representing individual menu items (e.g., File, Home, Insert).
    The user is using \emph{Wheel 1} with their \emph{index finger}. 
    By rotating the wheel, the user can move the focus in the menu.
    The other two wheels are used for navigating the second ((e.g., Font) and third-level menu items (e.g., Boldface, Italic).
    }
  \label{fig:teaser}
 \end{teaserfigure}

\maketitle
\section{Introduction}
\label{sec:intro}
Blind users rely on assistive technologies (AT) like screen readers to interact with application user interfaces (UIs).
On desktops, popular screen readers are NVDA~\cite{NVAccess38:online}, JAWS~\cite{jaws_screenshade}, VoiceOver~\cite{voiceover_osx}, which offer numerous keyboard shortcuts. 
A keyboard serves as the primary input device for blind users on desktops. 
Utilizing the operating systems' built-in accessibility support, commonly known as accessible APIs~\cite{msaa_arch, ui_automation, osx_api}, screen readers (SR) create a \T{DOM} tree-like textual meta-representation of an application~\cite{DOM2022}.
Navigating an app using SR and keyboard shortcuts is akin to navigating this underlying meta-representation. 
Most screen readers allow blind users to navigate either \emph{serially}, from left to right and top to bottom, using the \T{TAB} and directional \T{Arrow} keys, or \emph{hierarchically}, starting from the root node and proceeding to its children and grandchildren, using a combination of shortcuts. 
For example, users can press \T{Control+Option+Shift+Down} or \T{Up Arrow} to move into or out of a parent node in VoiceOver. 
However, both navigation strategies are slow and tedious, as blind users can go to only one of the four possible neighboring elements (left, right, up, down) at a time until they reach the target element.
Moreover, Recent studies have shown that apps requiring a higher average number of keystrokes for navigation are perceived as less accessible~\cite{touhidul2023probabilistic}.
For this reason, non-visual interaction can take up to three times longer than visual interaction~\cite{vi_difficulties, ram_prediction}.

To address the challenge of complex hierarchy navigation for blind users, we propose and design a mouse-shaped, three-wheel, stationary input device named \textsc{\textbf{\sysname{}}}.
The design of \sysname{} is informed by both prior works and based on the findings from multiple participatory design sessions with four blind individuals.
Figures~\ref{fig:teaser}a and ~\ref{fig:teaser}b show the \sysname{} prototype.
\sysname{}'s first mode of operation is the hierarchical navigation mode, or in short, the \textbf{\hnav{}} mode.
In \hnav{} mode, shown in Figure~\ref{fig:hnav_illustration}, the three wheels of the device are assigned to the top three levels of hierarchy in an app.
The user places three fingers on the three wheels and can rotate each wheel individually using a single finger or multiple fingers simultaneously.
The \hnav{} mode allows blind users to navigate complex multi-level UI hierarchies efficiently, using one wheel for each hierarchy level. 
There are also two buttons on the device that the user's thumb can access. 
The big button, which is larger, serves as the primary button, similar to the Left-click button of a mouse, and is expected to be used frequently. 
On the other hand, the small button serves as the secondary or Right-click button.
Additionally, \sysname{} has haptic feedback that notifies users when they have reached a boundary condition, such as the end of a list of menu items.

The ideation and design of \sysname{} draw inspiration from previous research that proposed rotary input devices to improve the speed of website and app navigation. 
Two such devices, Speed-Dial~\cite{Billah_speeddial} and NVMouse~\cite{haena-speeddial2}, were particularly influential.
In Speed-Dial~\cite{Billah_speeddial}, Billah et al. used a rotary input device (e.g., a Surface Dial~\cite{surface_dial}) to interact with web pages, demonstrating that blind participants performed data tasks significantly faster using the rotary input compared to screen-reader-provided keyboard-based navigation. 
Lee et al. reported similar findings with their NVMouse prototype~\cite{haena-speeddial1, haena-speeddial2}.
While the improvements by these approaches are noteworthy, 
Speed-Dial or NVMouse does not solve the difficulty of navigating user interface elements that belong to different sub-trees, such as c.2 and c.3 in Figure~\ref{fig:hnav_illustration}c. 
Users are still required to navigate through the parent nodes one by one until they reach the grandparent node that contains the target node.
\sysname{}'s three-wheel design overcomes this limitation by allowing blind users to navigate three different levels of hierarchy independently, with each wheel dedicated to a different level.

The three-wheel design of \sysname{} also offers versatility, allowing users to repurpose them for other tasks, 
such as 2D cursor manipulation, which is a feature that is available in certain screen readers such as JAWS. 
JAWS allows users to explore UI elements flatly, from left to right, top to bottom.
Moreover, our blind co-author in this paper mentioned using Windows Mouse Keys (MKs) to move the cursor when a keyboard alone is insufficient.
However, MKs are uncomfortable with long-distance cursor movement and lack precise speed control. 
Additionally, they do not offer helpful cursor localization feedback, only emitting beeps. 
To address this issue, we create a second mode for \sysname{}, the 2d navigation mode, or in short, the \textbf{\fnav{}} mode.
The illustration of \fnav{} mode is shown in Figure~\ref{fig:fnav_illustration}.
In this mode, the user uses \wa{} and \wb{} to move the cursor horizontally and vertically, respectively.
\wc{} is used to change the speed of the cursor movement.
At any time, the user can probe the cursor's location with respect to the top-left corner of the screen using the \texttt{CTRL} button.

A user study with $12$ blind participants showed that they needed 40\% less time doing hierarchy navigation tasks using \sysname{}'s \hnav{} mode compared to using the combination of keyboard and SR. 
In addition, they moved the cursor around the screen to acquire targets using \sysname{}'s \fnav{} mode, as instructed by a sighted confederate, enabling them to participate in mixed-ability collaborations.
This study, accompanied by a diary study with our blind co-author (Section~\ref{sec:diary}) also revealed that \sysname{} is easy to learn and easy to use; 
and it offers serendipitous benefits, 
such as collaborating with sighted users on a shared screen and clicking on partially accessible UI elements that are otherwise unreachable via a keyboard.

We summarize our contributions as follows:
\vspace{-\topsep}

\begin{itemize}
\item We design and develop a multi-wheel input device named \sysname{} that offers numerous benefits over typical keyboard-SR based non-visual interactions (Sections~\ref{sec:overview}, ~\ref{sec:design}, and ~\ref{sec:technical}).

\item \sysname{}'s \hnav{} mode allows blind users to navigate hierarchies in significantly less time than keyboard-SR-based navigation (Sections ~\ref{subsec:interaction-modes} and ~\ref{subsec:h-nav_eval}).

\item \sysname{}'s \fnav{} mode enables mixed-ability collaboration, allowing blind users to follow the commands issued by a sighted confederate and locate targets accurately in 2D space (Sections ~\ref{sec:tnav} and ~\ref{subsec:2d-nav_eval}).

\item Through a \new{diary study} with our blind co-author in this paper, we identify how \sysname{} provides solutions to unique problems, such as exploring graphical content without proper accessibility labels (Section~\ref{sec:diary}).

\end{itemize}
\section{Background and Related Work}
In this section, we first provide background on the inner workings of screen readers and the abstract UI tree.
Next, we position our work in the large literature on input devices and non-visual interaction.

\subsection{Construction of Abstract UI Tree for Screen Readers}
User interfaces (UIs) are typically designed with the assumption that the users have no perceptual and cognitive impairments and use a typical set of input and output devices~\cite{gajos2004supple}.
Thus, any mismatch between users' effective abilities and the underlying assumptions hampers the effectiveness of user interface design. 
Often, this diversity of needs is either ignored; or addressed via a manual redesign of the application UI; or via external assistive technologies (ATs), such as screen readers for blind computer users. 
Although a manual redesign is arguably the best~\cite{gajos2004supple}, it is neither feasible nor scalable because users' abilities and preferences vary, which can be hard to anticipate by the designers~\cite{bergman1995towards}. 

As a result, users' ability-specific adaptations are carried out by ATs. 
For example, screen readers adapt application UIs by creating a \emph{manifest} interface for users with vision impairments~\cite{about-face-2014}. 
These screen readers (e.g., NVDA~\cite{NVAccess38:online}, JAWS~\cite{jaws_screenshade}, and VoiceOver~\cite{voiceover_osx})  rely on the Operating System's accessibility support, a set of well-defined functions, commonly known as \textit{Accessibility APIs}~\cite{ms_automation, msaa_ui, osx_api, android_accessibility}, to extract a DOM-like hierarchical meta-representation of all UI elements in an app.
Each node in this tree contains a textual meta-representation (e.g., name, states) of the UI element it represents.
This DOM-like abstract UI tree is invisible to sighted users but accessible via screen readers' keyboard shortcuts---when a blind user selects a node in this tree, the screen reader reads out the textual description of the corresponding UI element (e.g., \texttt{\small "OK Button"}) loudly.
In short, the abstract UI tree and screen readers compensate for the inability of users to see a graphical user interface and implement keyboard shortcuts for users as an alternative to using a mouse to point-and-click graphical elements.

Unfortunately, these adaptations come at the expense of user experience. 
Prior work~\cite{harris_interoperability} has shown that when the output modality of a UI-rich application is manifested from its original form to another modality, i.e., consuming an application aurally instead of visually, the adaptation introduces undesired side effects, such as the application may become partially accessible and the task completion time may increase rapidly. 
For instance, Billah et al.~\cite{Billah_speeddial} reported that filling out an online form on a travel-booking website could take $224s$ for blind users. On the other hand, the sighted authors of this paper performed a similar task for less than $60s$, which indicates that the task completion time for blind users, in this case, is 3$\times$ more than that of a sighted individual.

\subsection{Pointing Devices for Interaction}
Pointing devices, such as computer mouses (or mice), were first conceptualized in the sixties~\cite{engelbart1962augmenting, english1967display} and had become an effective input device to interact with graphical user interfaces (UIs) on two-dimensional (2D) screens. 
In early prototypes, users needed to move the mouse pointer (i.e., \textit{cursor}) along $X$ and $Y$ axes on the screen by rotating a pair of wheels.
These wheels were later replaced by buttons~\cite{sherr2012input}  in optical mice~\cite{popov2004laser}. 
The current generation of mice still have a single wheel but for a different purpose, scrolling~\cite{hinckley2002input}.  
Pointing devices are commonly evaluated by two metrics: \emph{performance} and \emph{comfort}~\cite{douglas1999testing}.
Below, we describe these two metrics.

\subsubsection{Performance Measure of Pointing Devices}
Fitts' law~\cite{fitts1954information} is commonly used to measure the performance of pointing devices, which include several metrics, such as movement time (i.e., the time required to get to a UI object from another), error rate (i.e., \% of mistakes made during a specific task), and throughput (a metric combining both movement time and accuracy).
Fitts' law states that when moving a cursor from a source UI to a destination UI, the index of difficulty (\textbf{ID}, in bits), or in short, the difficulty, is proportionate to the distance between the source and the destination and inversely proportional to the width (or area) of the destination. 
Its initial formulation was for 1D, which Mackenzie et al.~\cite{mackenzie1992extending} extended to 2D by replacing the width with the target's height if the height is smaller or calculating the width along the direction of approach to the target. 
For UI elements having the same height and width (e.g., icons), considering only the width of the target is sufficient. 
Mackenzie et al.~\cite{mackenzie2001accuracy} also propose two extended static measures, orthogonal direction change, and movement direction change; two dynamic measures, movement variability, and movement error, to make the comparisons more comprehensive for individuals with no vision impairments. 
In our evaluation, we found that static measures are relevant to evaluate \sysname{}'s performance, especially in \fnav{} mode with staircase-like 2D movements.

\subsubsection{Comfort of Pointing Devices}
The comfort of pointing devices is subjective and measured by asking questions on rating different aspects, such as users' physical effort, fatigue and comfort (e.g., hand/wrist posture comfort, clicking comfort), speed and accuracy, and overall usability~\cite{douglas1999testing, dehghan2015assessment, coelho2017psychometric}.
However, having too many questions can be confusing to users. Therefore, we evaluated \sysname{} on three aspects: clicking comfort, satisfaction,  and overall usability. 

\subsection{Adaptation of Pointing Devices for Non-Visual Interaction}

Pointing devices like mice are hardly used by blind users because the pointer (or the mouse cursor) only provides \textit{visual} feedback~\cite{Billah_speeddial}, which blind users cannot perceive.
Therefore, researchers attempted to replace this visual feedback with an alternative, such as 2D translational force, vibration, skin stretch, and thermal feedback to convey kinesthetic and tactile information regarding the object under the cursor~\cite{haptic_mouse_1, haptic_mouse_3}.
Often, this alternative feedback is presented on a small tactile display on one side of the mouse.
However, the spatial resolution of this display is coarse due to the small display size. 
Kim et al.~\cite{kim2008inflatable} designed an inflatable mouse that uses an air balloon, which senses the pressure from users' hands and thus acts as an input device.
Nevertheless, this device is tiring for tasks requiring high finger pressure and precision.

A large body of literature on haptic feedback is devoted to Braille displays, auditory feedback, and tactile pin arrays~\cite{afb_braille_displays, assistive_haptic_feedback, vi_aug_geomagic, moose, habos, vi_multimodal, pin_array, tactile_web_browser, haptic_glove, soviak2015feel, soviak2015haptic}.
Some of the work proposed guidelines for designing appropriate haptic sensations for blind users~\cite{assistive_haptic_feedback}. 
Soviak et al.~\cite{haptic_glove, soviak2015feel, soviak2015haptic} explored an alternative, glove-like input device to convey UI elements' boundaries via audio-haptic feedback.
These prototypes demonstrated the benefits of haptic feedback for non-visual interaction in obtaining an overview of a web page; however, the unrestrained freedom of movement within the page hinders precise navigation and information finding. 
Unlike the above prototypes, \sysname{} uses haptic feedback to inform user actions and announce screen or UI boundaries only.

\begin{table*}[!t]
    \centering
    \begin{tabular}{|C{0.5cm}|C{2cm}|C{3cm}|C{7.5cm}|}
        \hline
        \textbf{\textsc{No.}} & \textbf{\textsc{Mode}} & \textbf{\textsc{Input/Action}} & \textbf{\textsc{Outcome}} \\
        \hline
        \cellcolor{gray!10}1 & \multirow{3}{*}[-0.75cm]{\hnav{}} & \cellcolor{gray!10}\wa{} scroll & \cellcolor{gray!10}Navigate hierarchy level currently mapped to \wa{} \\
        \cline{1-1}\cline{3-4}
        2 &  & \wb{} scroll & Navigate hierarchy level currently mapped to \wb{} \\
        \cline{1-1}\cline{3-4}
        \cellcolor{gray!10}3 &  & \cellcolor{gray!10}\wc{} scroll & \cellcolor{gray!10}Navigate hierarchy level currently mapped to \wc{} \\
        \cline{1-1}\cline{3-4}
        4 &  & \texttt{CTRL} + Primary button press & Move all three wheel's assignments one level down in the app hierarchy \\
        \cline{1-1}\cline{3-4}
        \cellcolor{gray!10}5 &  & \cellcolor{gray!10}\texttt{CTRL} + Secondary button press & \cellcolor{gray!10}Move all three wheel's assignments one level up in the app hierarchy \\
        \hline
        6 & \multirow{4}{*}[-0.4cm]{\fnav{}} & \wa{} scroll & Move cursor horizontally \\
        \cline{1-1}\cline{3-4}
        \cellcolor{gray!10}7 &  & \cellcolor{gray!10}\wb{} scroll & \cellcolor{gray!10}Move cursor vertically \\
        \cline{1-1}\cline{3-4}
        8 &  & \wc{} scroll & Adjust cursor speed \\
        \cline{1-1}\cline{3-4}
        \cellcolor{gray!10}9 &  & \cellcolor{gray!10}Secondary button press and hold for 300ms & \cellcolor{gray!10}Turn on or off \tnav{} mode \\
        \cline{1-1}\cline{3-4}
        10 &  & \texttt{CTRL} press & Announce the cursor's location on the screen \\
        \hline
        \cellcolor{gray!10}11 & \multirow{3}{*}[-0.2cm]{\hnav{} or \fnav{}} & \cellcolor{gray!10}Primary button press & \cellcolor{gray!10}Simulate a mouse left click \\
        \cline{1-1}\cline{3-4}
        12 &  & Secondary button press & Simulate a mouse right click \\
        \cline{1-1}\cline{3-4}
        \cellcolor{gray!10}13 &  & \cellcolor{gray!10}\texttt{CTRL} + Both Button Press & \cellcolor{gray!10}Switch between \hnav{} and \fnav{} modes \\
        \hline
    \end{tabular}
    \caption{Summary of \sysname{}'s inputs and outcomes in different modes.}
    \label{tab:inputs}
\end{table*}

\subsection{Rotary Input for Interaction}
Another line of work represents UI elements on the two-dimensional (2D) screen into a one-dimensional (1D) circular list, then uses rotary devices (e.g., dials, wheels) to rapidly navigate through the list~\cite{Billah_speeddial, haena-speeddial1, haena-speeddial2}. 
Because rotary input (e.g., rotate clockwise/anti-clockwise) is inherently one-dimensional, this mode of interaction is more fitting and efficient for blind users when navigating a list.
In fact, sighted users also benefit from rotary input with a mouse wheel for scrolling~\cite{hinckley2002input}.
Lee et al.~\cite{lee2010design} extended the rotary input for sighted users from a single mouse wheel to three wheels by placing two virtual wheels on both sides of the physical wheel. 
Their experiment revealed that users performed certain operations (e.g., volume up or volume down in a media player) up to two times faster than a single wheel. In addition, the users get better with more iterations of the same tasks.
These encouraging findings inspired us to design \sysname{} with multi-wheels.

\subsection{Accommodations for Mouse and Virtual Cursors}
\label{sec:mousekeys}
Operating Systems (OSes) often support accommodations for mouse-based interaction. 
For example, Windows OS allows users to move the mouse cursor by pressing Num keys~\cite{windows_mouse}, commonly known as Mouse Keys; 
MacOS supports controlling the native VoiceOver~\cite{voiceover} screen reader through trackpad inputs~\cite{voiceover_touchpad}. 
However, during our participatory design sessions, we found that blind participants rarely use these accommodations due to usability issues.   

Screen readers usually provide multiple cursors. For example, JAWS~\cite{jaws_screenshade} screen reader offers three cursors: \textit{PC cursor}, which is the caret in Word/text documents; \textit{JAWS cursor}, which is the mouse pointer; and \textit{virtual PC cursor}, which simulates an invisible caret in webpages.
NVDA~\cite{nvda} also supports similar cursors (e.g., NVDA's \textit{review cursor} is similar to JAWS's \textit{virtual PC cursor}).
In addition, screen readers can be configured to report the textual content directly beneath the mouse cursor as the user moves the cursor~\cite{nvda_mouse}. 
Multiple cursors are useful for different applications and tasks, as indicated by our participants.
\sysname{} design aligns with this notion of multiple cursors.

\section{Overview of \sysname{}}
\label{sec:overview}

\sysname{} is a mouse-shaped input device with three wheels and two push buttons on the side, as shown in Figure~\ref{fig:teaser}. 
All three wheels are identical in size. 
However, the primary button is slightly bigger than the secondary one.
The bigger button's role is similar to that of the left click or the primary button of a typical mouse. 
Likewise, the smaller button acts as the right-click button of a mouse.
A user can grip the device with their right hand so that their index finger rests on the first wheel, 
the middle finger on the second wheel, the ring finger on the third wheel, and the thumb over the two buttons, as shown in Figure~\ref{fig:teaser}.
\new{When assembled, \sysname{} measures 103 mm in length, 60 mm in width, and 33mm in height, meeting the ANSI standard range~\cite{american2007ansi} for mouse dimensions: 120 mm in length, 40-70 mm in width, and 25-40 mm in height.}

Unlike a mouse, \sysname{} is stationary, i.e.,  users do not move it on the surface when using it. 
Instead, they rotate wheels to maneuver the cursor.
In our current design, \sysname{} is connected to a computer via a USB cable. 
However, a Bluetooth-based wireless connection is feasible. 

\sysname{} provides various audio-haptic feedback to communicate the current context of the cursor. 
\sysname{} has a buzzer and haptic motor on its motherboard. 
The buzzer emits a beep during significant events, while the haptic motor creates a gentle vibration with each rotation---none interfere with the audio output from screen readers.

\sysname{} primarily operates in two modes: (i) a hierarchical navigation (\hnav{}) mode and  (ii) a two-dimensional or flat navigation (\fnav{}) mode. 
Based on the feedback from a diary study with our blind co-author (Section~\ref{sec:diary}), 
we later introduced another mode named \tnav{}, which is a special case of \fnav{} mode facilitating the teleportation of the mouse cursor.

Table~\ref{tab:inputs} provides an overview of various input methods and their corresponding results when using \sysname{}, considering the device's current operating mode.
It is worth noting that we have strategically incorporated the keyboard \texttt{CTRL} button as an input modifier in three different scenarios. 
This approach simplifies user interaction by requiring them to remember just one key on the keyboard.

\subsection{Interaction Using \sysname{}: Hierarchical Navigation (\hnav{} Mode)}
\label{subsec:interaction-modes}

\begin{figure*}[t!]
  \centering
  \includegraphics[width=.99\linewidth]{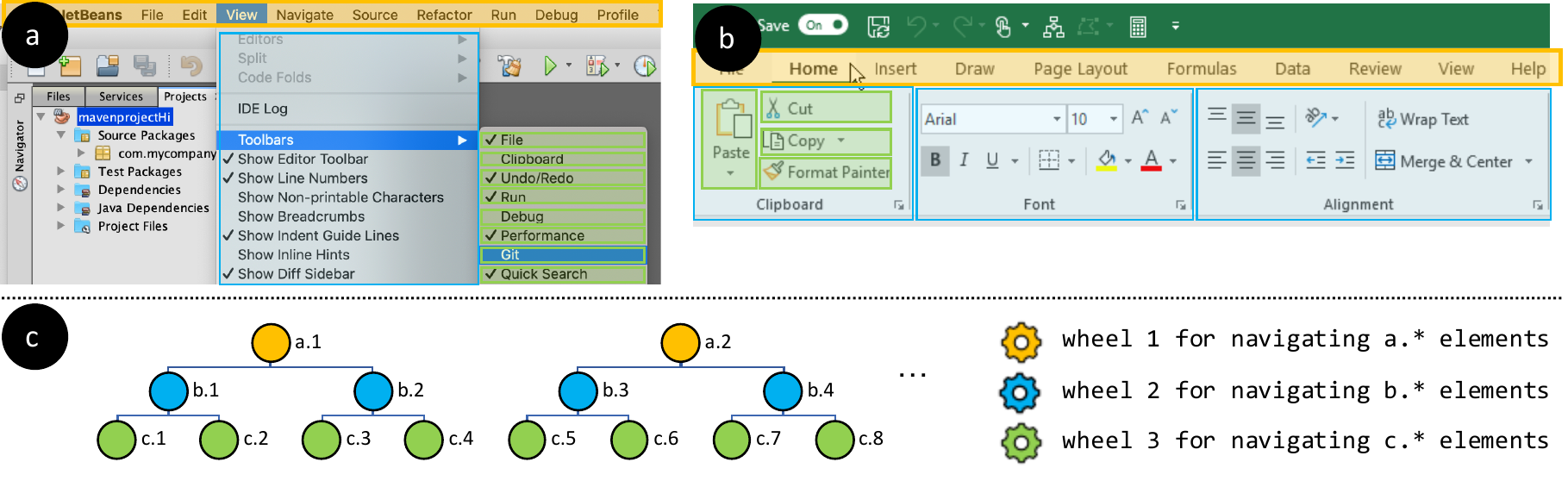}
  \caption{
  Demonstration of \sysname{}'s \hnav{} mode. (a) Multi-level menus in NetBeans; (b) nested structures in ribbons in MS Word; (c) a sample 3-level tree hierarchy to represent menus in either (a) or (b). For example, $1^{st}$-level elements, \texttt{\small \{a.1, a.2, ...\}} could represent the top-level menu items like \texttt{\small \{File, Edit, View, ... Profile\}} in (a) or \texttt{\small \{Home, Insert, ..., Help\}} in (b). 
  Assuming \texttt{\small a.1} equals \texttt{\small Home}, the $2^{nd}$-level menu items, \texttt{\small \{b.1, b.2, ...\}} will be \texttt{\small \{Clipboard, Font, ...\}} in (b).
  Similarly, assuming \texttt{\small b.1} equals \texttt{\small Clipboard}, the $3^{rd}$-level menu items, \texttt{\small \{c.1, c.2, ...\}} will be \texttt{\small \{Paste, Cut, ...\}} in (b). 
  In \hnav{} mode, \wa{} is always mapped to $1^{st}$-level menu items (i.e., \texttt{\small a.*}), \wb{} is mapped to $2^{nd}$-level menu items (i.e., \texttt{\small b.*}), and \wc{} is mapped to $3^{rd}$-level menu items (i.e., \texttt{\small c.*}).}
  \label{fig:hnav_illustration}
\end{figure*}

In \hnav{} mode, \sysname{} navigates the abstract UI tree of an app, as shown in Figure~\ref{fig:hnav_illustration}.
By default, three wheels of \sysname{} point to the top three levels in an app's DOM.
Each wheel maintains its own cursor---making a total of three independent cursors. 
Further, each wheel maintains its own state. 
For example, a wheel remembers the last UI object a user had focused on the last time in a level and resumes interacting from that element when the user focuses back.
Thus, it eliminates the need to explore elements from the beginning in a hierarchy. 

The rotate action (e.g., clockwise or counterclockwise) selects an element bidirectionally at a level. 
While \texttt{wheel 1} can select any element in the $1^{st}$ level, \texttt{wheel 2} only selects the immediate children of the element currently selected by \texttt{wheel 1}. 
Recursively, \texttt{wheel 3} only selects the immediate children of the element selected by \texttt{wheel 2}.
When \wa{}'s cursor moves to a certain node in the UI tree, 
\wb{}'s cursor automatically moves to the first child of the node selected by \wa{}.
Similarly, \wc{}'c cursor automatically moves to the first child of the node selected by \wb{}.
Figure~\ref{fig:hnav_illustration}c demonstrates how the menus and ribbons of two applications can be organized in a tree hierarchy and mapped in \sysname{}.

To perform a left-/right-click operation, the user presses the primary/secondary side buttons.
Users can define the rotation resolution (in degrees) to adjust a wheel's sensitivity. 
Upon rotation, \sysname{} provides audio-haptic feedback to affirm a valid operation and sometimes to convey spatial information, such as whether a UI element is the last (or first) among its siblings.

\subsubsection{\hnav{} Mode vs. Using Keyboard and Screen Reader}

\begin{figure}[!ht] 
    \centering
    \includegraphics[width=\columnwidth]{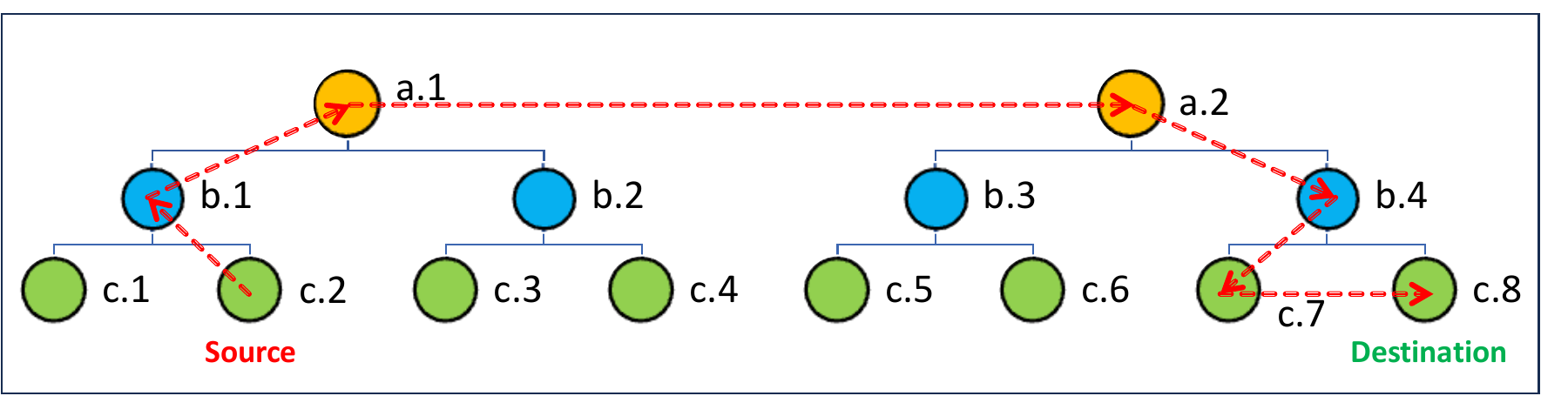}\\
    \caption{The red arrows highlight the path (\texttt{\small c.2} -> \texttt{\small b.1} -> \texttt{\small a.1} -> \texttt{\small a.2} -> \texttt{\small b.4} -> \texttt{\small c.7} -> \texttt{\small c.8}) a blind user would have to take using a combination of keyboard and screen reader when going from \texttt{\small c.2} to \texttt{\small c.8} in the hierarchy from Fig.~\ref{fig:hnav_illustration}c.
    There are \textbf{six} steps in total, and at least six keystrokes would be required.
    }
    \label{fig:kb_sr}
\end{figure}

\begin{figure*}[!ht]
\centering
    \begin{minipage}[t]{0.49\linewidth}
        \centering
        \includegraphics[width=\columnwidth]{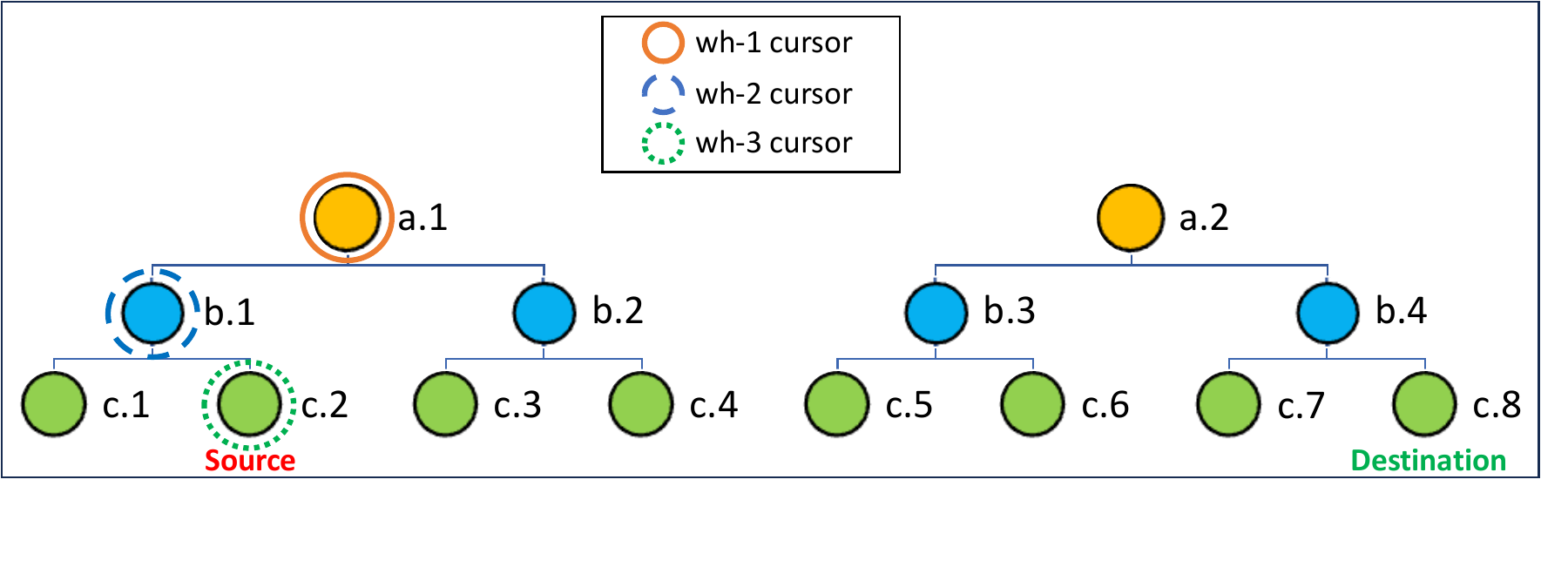}\\
        \small(a)~The cursors for the three wheels at their initial position.
    \end{minipage}
    \hfill
    \begin{minipage}[t]{0.49\linewidth}
        \centering
        \includegraphics[width=\columnwidth]{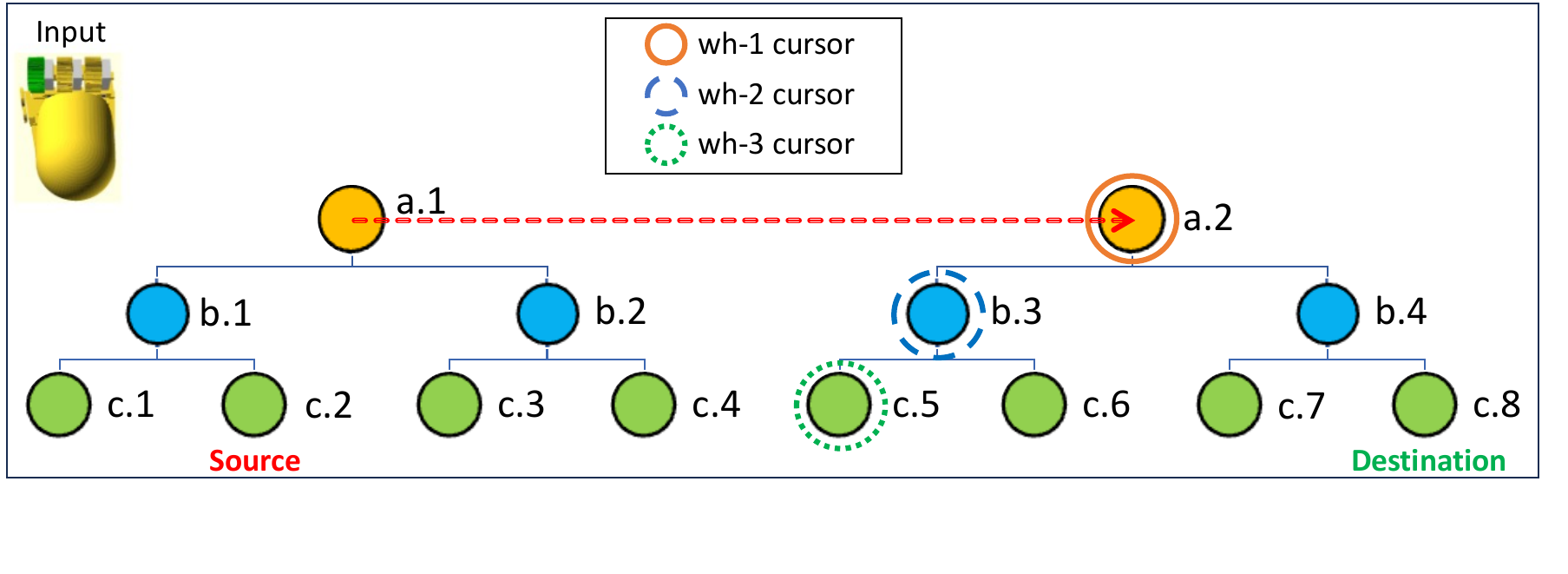}\\
        \small(b)~User rotates \wa{} to move its cursor from \texttt{\small a.1} to \texttt{\small a.2}. The cursors for \wb{} and \wc{} automatically move to \texttt{\small b.3} and \texttt{\small c.5}, respectively.
    \end{minipage}
    
    \medskip
    
    \begin{minipage}[t]{0.49\linewidth}
        \centering
        \includegraphics[width=\columnwidth]{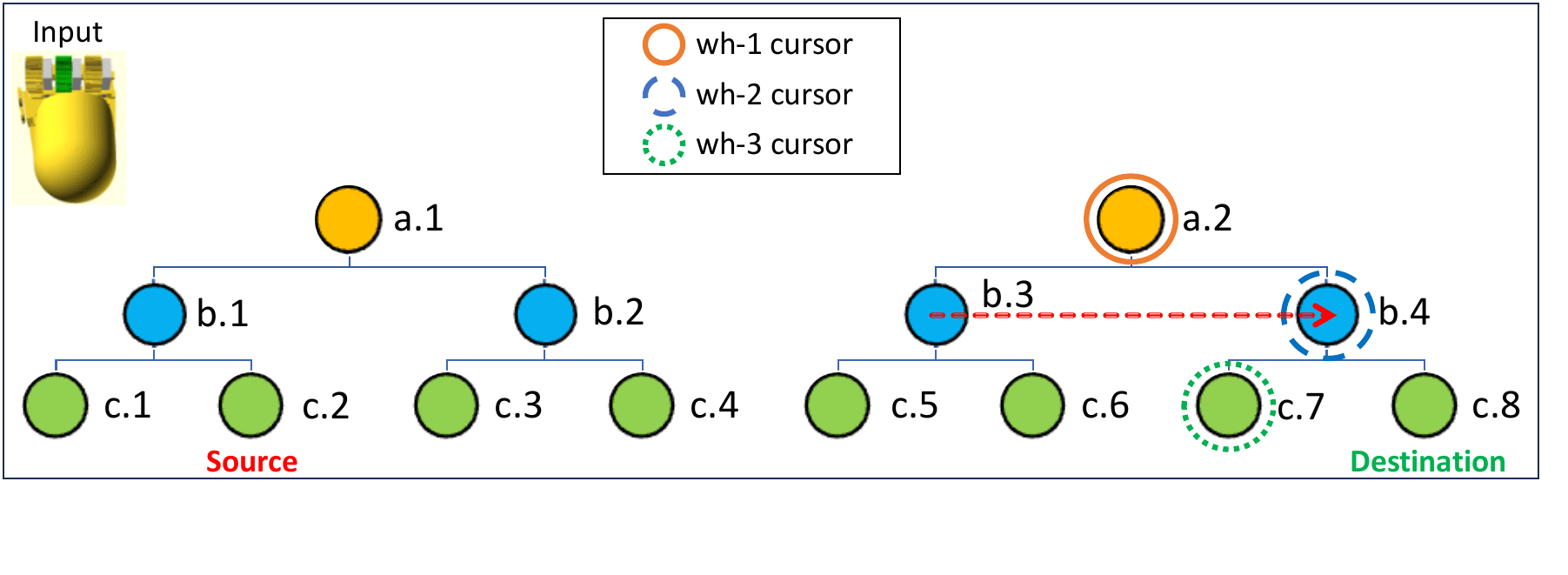}\\
        \small(c)~User rotates \wb{} to move its cursor from \texttt{\small b.3} to \texttt{\small b.4}. The cursor for \wc{} automatically moves to \texttt{\small c.7}.
    \end{minipage}
    \hfill
    \begin{minipage}[t]{0.49\linewidth}
        \centering
        \includegraphics[width=\columnwidth]{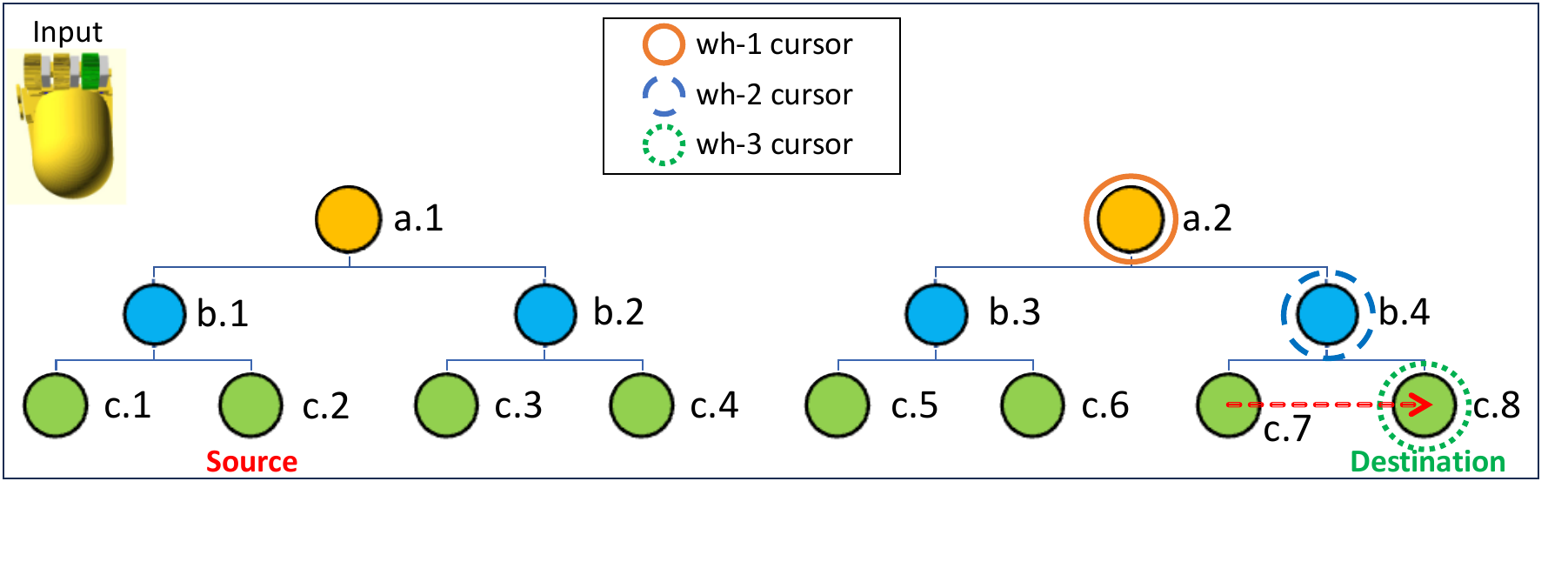}\\
        \small(d)~User rotates \wc{} to move its cursor from \texttt{\small c.7} to \texttt{\small c.8}
    \end{minipage}

    \caption{Navigating from
    \texttt{\small c.2} to \texttt{\small c.8} in the hierarchy from Fig.~\ref{fig:hnav_illustration}c using \sysname{}'s \hnav{} mode.
    (a)-(d) shows the positions of the three \sysname{} cursors mapped to its three wheels at different stages of the navigation.
    Only \textbf{three} rotations are required in total.}
    \label{fig:wheeler-h-nav-steps}
\end{figure*}

To demonstrate the advantage of \hnav{} mode over using a  keyboard and a screen reader combo,
suppose a user wants to move from node \texttt{\small c.2} (source) to node \texttt{\small c.8} (destination) in Figure~\ref{fig:hnav_illustration}c.

Fig.~\ref{fig:kb_sr} shows the navigation steps when the user is using a keyboard and a screen reader.
Note that it would require at least six operations (i.e., \texttt{TAB} or \texttt{[SHIFT+TAB]} key press) in total.

Fig.~\ref{fig:wheeler-h-nav-steps}a-~\ref{fig:wheeler-h-nav-steps}d shows the navigation steps when the user is using \sysname{} for the same navigation task.
Notice how the user can complete the task in three rotations.

\subsubsection{Traversing Apps with More than 3 Levels.}
If an application has more than 3 levels, the user can move all three cursors one level down in the app hierarchy by pressing \sysname{}'s Primary button while holding the \texttt{CTRL} key.
Likewise, to move all three cursors upward, they can do so by holding the \texttt{CTRL} key and pressing \sysname{}'s Secondary button.
These actions are also shown in Table~\ref{tab:inputs}.

\subsection{Interaction Using \sysname{}: Two-Dimensional  Navigation (\fnav{} and \tnav{} Modes)}
\label{sec:tnav}

\begin{figure*}[t!]
\centering
  \includegraphics[width=.75\linewidth]{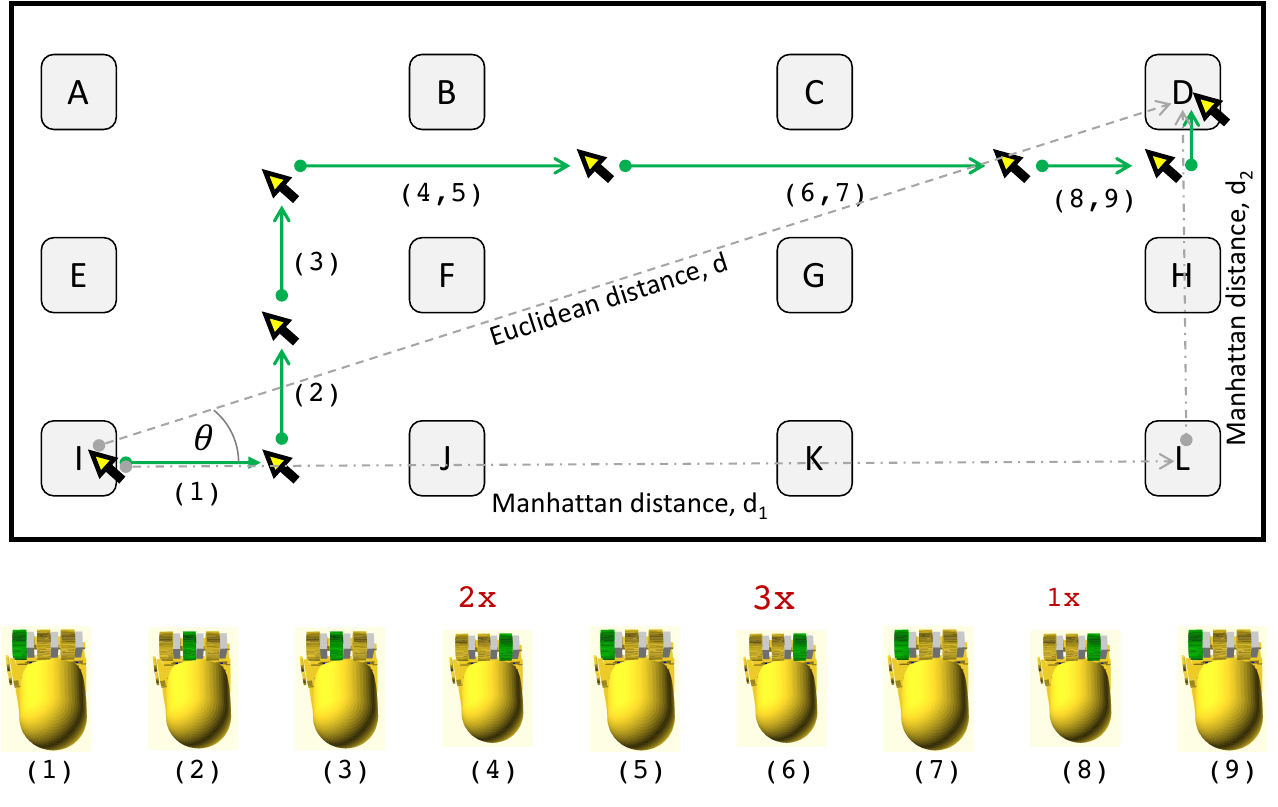} 
    \caption{Illustrations of \sysname{}'s \fnav{} mode. (Top) A blind user moves the cursor from the lower-left corner to the upper-right corner of a 2D screen with 12 UIs organized in a $3\times4$ grid. (Bottom) shows the sequence of wheel operations the user performs at different steps to achieve this goal. Each step \texttt{(i)} is marked in both the top and the bottom figures. The green-colored wheel indicates which wheel the user rotated at step \texttt{(i)}. The user rotates \wa{} or \wb{} to move the cursor horizontally or vertically and adjusts the speed of the cursor by rotating \wc{}. Note that the Euclidean distance between the source and the destination is $d$, which sighted users usually take using a mouse. In contrast, blind users take the Manhattan distance between two nodes (e.g., $d_1$ along X, plus $d_2$ along Y) in \fnav{} mode.} 
    \label{fig:fnav_illustration}
\end{figure*}

In \fnav{} mode, the wheels have different roles: 
\wa{} moves the cursor along the \texttt{X-axis}, \wb{} moves it along the \texttt{Y-axis}, and 
\wc{}  is used to control the speed of the cursor movement. 
\fnav{} mode is demonstrated in Figure~\ref{fig:fnav_illustration}, which depicts a scenario of a blind user moving the cursor on a 2D screen from the lower-left corner to the upper-right corner.
While moving the cursor, the user can rotate \wc{} to change the speed of the cursor movement.
\new{Users can scroll both \wa{} and \wb{} simultaneously and it would result in diagonal cursor movement.}

Loss of context is a prominent issue for users with visual impairment while navigating in 2d space~\cite{islam2023spacex}.
To address this, users can probe the cursor location anytime by pressing the \texttt{CTRL} key in \fnav{} mode. 
When \texttt{CTRL} is pressed, \sysname{} reads out the cursor location as a percentage of \texttt{X} and \texttt{Y} coordinates with respect to the screen width and height. 
For instance, if the user's cursor is above item \texttt{B} in Figure~\ref{fig:fnav_illustration}, \sysname{} would read out something like \emph{``30\% from the left and 10\% from the top''}.
On cursor-hover, \wh{}'s built-in TTS engine automatically reads out the name of a UI element. 

\subsubsection{\tnav{} Mode.}
\tnav{} is a special case of \fnav{} mode, where \sysname{} teleports the mouse cursor to the closest neighboring UI along the direction of the cursor movement. 
This is faster than \fnav{} to move from one element to another.

\subsection{Toggling Modes}
To toggle between \hnav{} and \fnav{} modes, users have to hold the \texttt{CTRL} button and press both the primary and secondary buttons of \sysname{} at the same time.
If \sysname{} is in \fnav{} mode, the users can turn on or off the \tnav{} mode by pressing and holding the secondary  (i.e., the small) button for some time (e.g., 300 ms).

\section{Ideation of \sysname{}}
\label{sec:design}

The initial design of \sysname{} is informed by the literature.
Below, we synthesize relevant prior work and describe how it inspired our design.

\subsection{Three Rotary inputs}
We observed that certain desktop screen readers (SRs), such as VoiceOver and ChromeVox, allow users to navigate application UIs hierarchically, similar to traversing an HTML DOM tree. 
These SRs let users go to a UI's immediate parent, child, or siblings one at a time. 
Often, UI elements next to each other visually belong to different parents, i.e., sub-trees, in the DOM.
It creates a challenge for blind users because they must traverse different sub-trees to navigate those elements. 
This observation led us to design an input device that lets blind users traverse different sub-trees independently.

We were also inspired by Speed-Dial~\cite{Billah_speeddial}, where Billah et al. demonstrated that a rotary input device could emulate mouse-like functions for blind users. 
However, Speed-Dial does not address the challenge of navigating UI elements in different sub-trees---users still need to go to the parent nodes individually until they find the grandparent whose child is the target node. 
Therefore, we conceptualized a hypothetical device with three rotary inputs, where each rotor is mapped to a level to reduce the number of times users need to go up in the parents.
\new{Moreover, developers often organize their apps hierarchically using a standard template---at the high level, there are menus, toolbars, sidebars, status bars, and client areas, each of which can have a second level, e.g., sub-menus, split toolbars, and containers/groups; and the most interactive elements (e.g., buttons, text areas) appear at the third level. This template inspired us to use three wheels, one for each level, to maximize coverage.}

\new{However, we found that the UI hierarchy of most applications spans more than three levels. 
We considered adding another wheel (under the pinky), but in our design mock-up, we found rotating this wheel difficult. 
This is due to the connection between the intrinsic muscles of human hands and innervation---the pinky and medial half of the ring finger are connected with the ulnar nerve (for gross hand movement). 
In contrast, the index, middle, and lateral half of the ring finger are connected by the median nerve (for more precise hand movement). 
As such, we kept the number of wheels to three, assigning the most frequently used wheel under the index and the least frequently used wheel under the ring.}

\new{For a similar reason, we placed the wheels vertically so that a finger only requires flexion/extension movement, for which maximum biomechanical advantage is available (e.g., more muscle groups are involved). 
Placing the wheel horizontally will require a finger abduction/adduction movement, for which fewer muscle groups are involved.}

\subsection{Flat 2D Navigation Mode}
We also observed that certain screen readers (e.g., JAWS) allow users to explore UI elements flatly, from left to right, top to bottom.
Further, the blind co-author of this paper stated that he occasionally uses  Windows  Mouse Keys (MKs) to manipulate the cursor when an element is not reachable by the keyboard. However, MKs are hard to use--- pressing these arrow keys for long-distance cursor movement is uncomfortable.
In addition, MKs lack a fine-grained cursor pace control.
Moreover, MKs do not provide useful feedback to localize the cursor; it only beeps. 
These pieces of information motivated us to devise a separate mode in our device that can act as a usable mouse for blind users.

\begin{figure}[t!]
  \centering
  \includegraphics[width=.9\columnwidth]{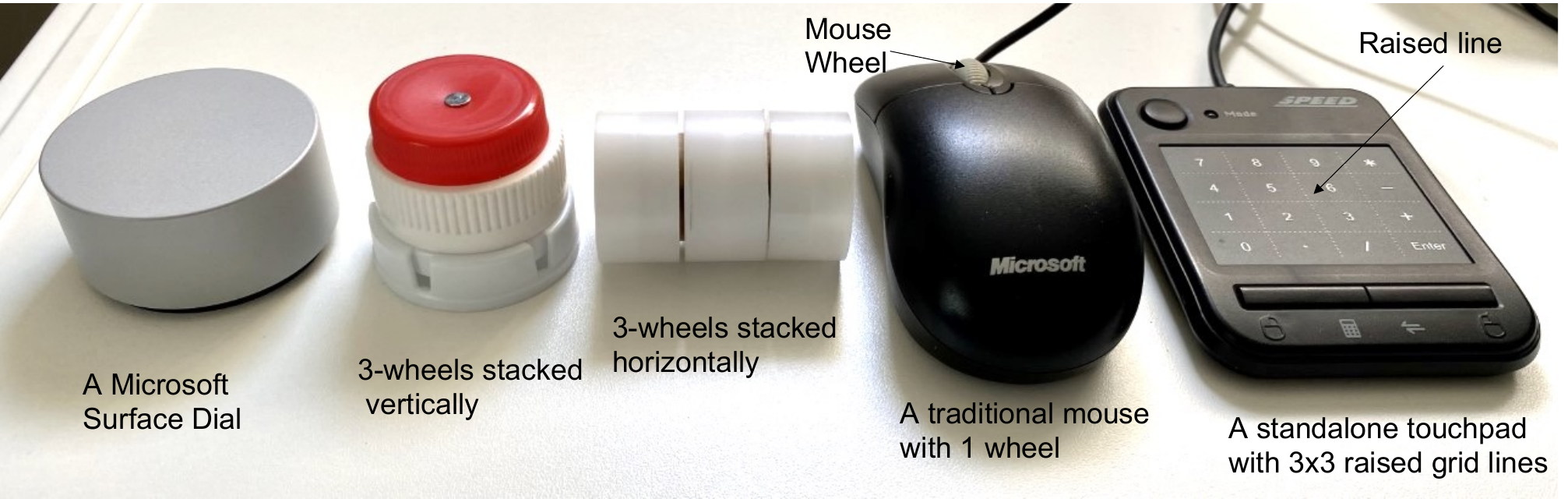}
  \caption{Several low-fidelity prototypes during ideation.}
  \label{fig:prototype_lowfi}
\end{figure}

\subsection{Initial Design}
After conceptualizing the design, we organized a participatory design session with four blind individuals.
The session aimed to brainstorm the design of a device capable of embodying our navigational paradigm.
We presented several low-fidelity prototypes (shown in Figure~\ref{fig:prototype_lowfi}) with a different number of rotary inputs as well as existing input devices that sighted users use.
These include a Surface Dial, a 3-wheel vertical device, a 3-wheel horizontal device, a mouse, and a touchpad that supports multi-touch.

We explained our design idea to them and asked their opinion about a usable device. 
\Partis{} preferred the form factor of the Surface Dial 
and mentioned that a 3-wheel vertical device could be an extension to Surface Dial but reported difficulty in rotating multiple wheels all at the same time.
All \partis{} preferred the 3-wheel horizontal device because they could comfortably place 3 fingers over 3 wheels. 
However, they mentioned that rotating the wheels without supporting their palm was difficult. 
When one participant pointed out the idea of combining a mouse with the 3-wheel horizontal device, all \partis{} were elated; they remarked that it could be a viable prototype.
When asked about the touchpad supporting multi-finger inputs, everyone was firmly against it. 
They contended that employing three fingers simultaneously would lead to a highly counter-intuitive experience, as moving one finger in one direction and the others in another direction would be confusing. 
Essentially, they believed they had to swipe all three fingers either upward or downward, which did not align with our intended concept of three \emph{independent} cursors.

Once we established the initial design, we asked the participants where to place the two mouse buttons. 
Three participants proposed making both buttons easily accessible by the thumb, suggesting placing them on the left side. 
Other participants initially recommended placing one button on each side, but they soon recognized that pressing the right button with the pinky finger would be challenging. 
In conclusion, the most promising design that emerged was a 3-wheel horizontal device resembling a mouse, 
featuring two buttons that were easily accessible by the thumb.

\subsection{Design Iterations}
\label{subsec:design_iterations}
The version of \sysname{} we presented in this paper resulted from three major design iterations. 
As one of the authors of this paper is a blind power user, we had the privilege to discuss, update, and evaluate mini-iterations of \sysname{} in-house. 
In each major iteration, we invited the same four blind individuals to collect their feedback and recommendations and to ensure that we incorporated their earlier feedback into the current iteration of the design.
Appendix~\ref{appendix:design_iterations} contains more details on these design iterations.

\section{\sysname{} Implementation}
\label{sec:technical}
This section discusses \sysname{}'s hardware implementation, electrical components, and firmware design.

\begin{figure}[t!]
    \centering
        \centering
        \includegraphics[width=0.8\columnwidth]{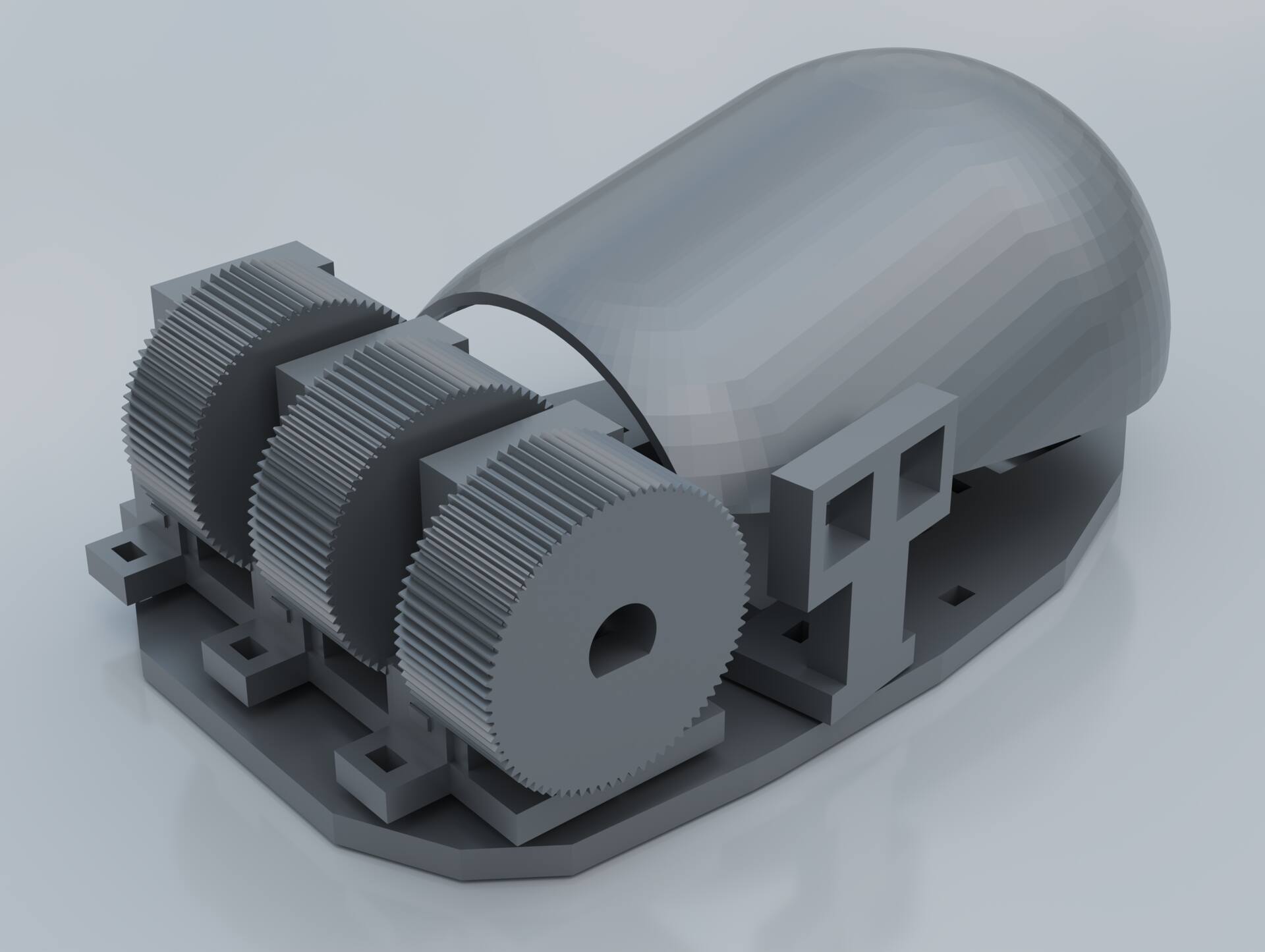}
        \caption{A high-level rendering of \sysname{} putting individual 3D parts together.}
        \label{fig:device3}
\end{figure}


\begin{figure}[t!]

    \centering
    \includegraphics[width=1\columnwidth]{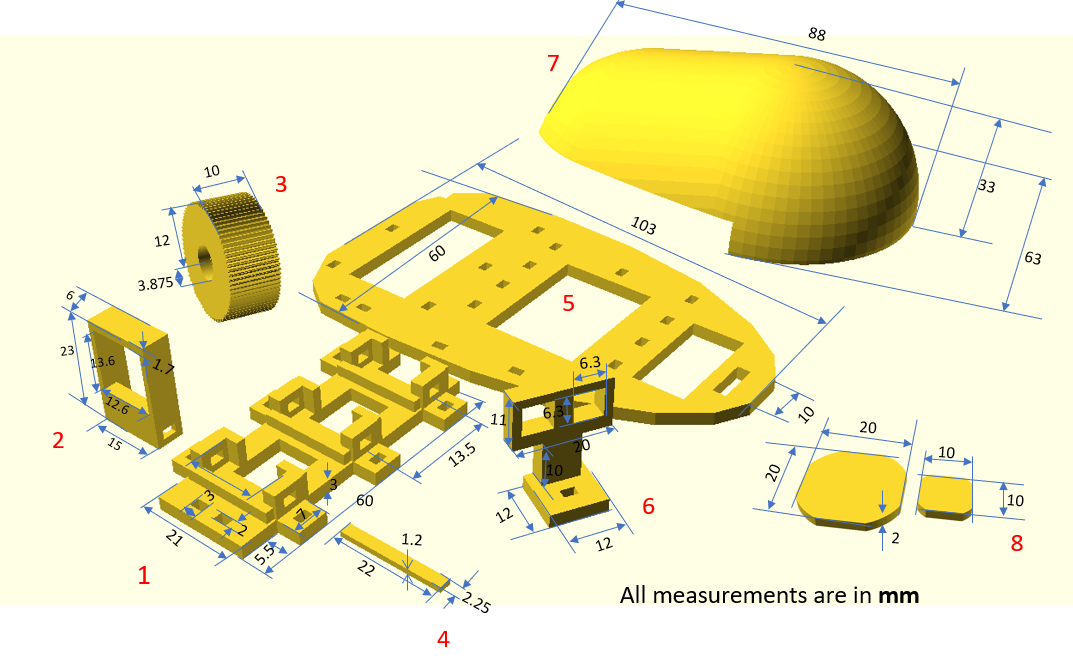}
    \caption{Individual 3D components of \sysname{} with dimensions in millimeters: (i) wheelbase, (ii) encoder holder, (iii) wheel, (iv) connector, (v) main base, (vi) button stand, (vii) top shell, and (viii) button covers.
    }
    \label{fig:3d-parts}

\end{figure}

\begin{figure}[t!]
    \centering
    \begin{minipage}{.49\textwidth}
        \centering
        \includegraphics[width=1\columnwidth]{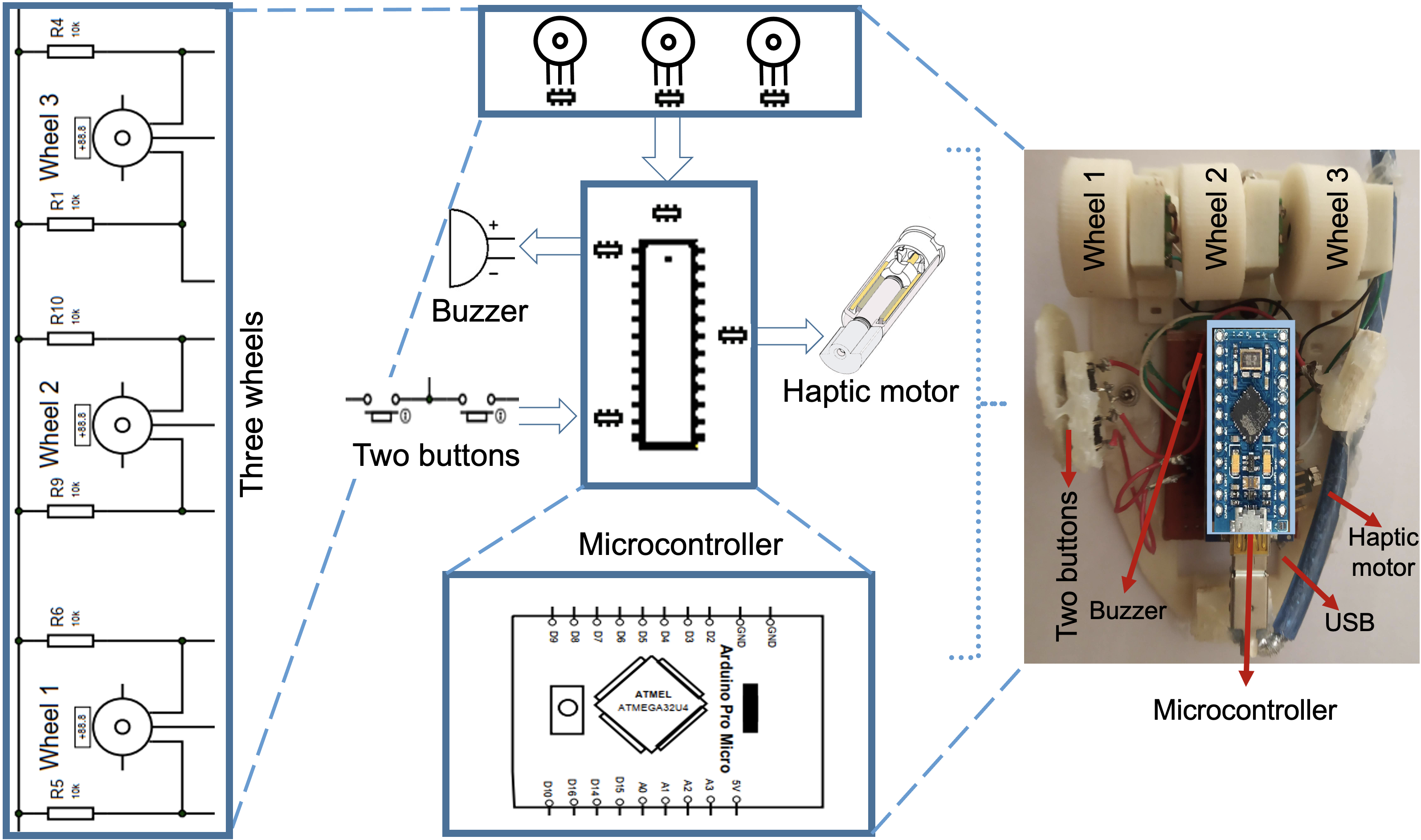}
        \caption{A high-level block diagram of the major electrical components of \sysname{}. The microcontroller measures the rotation of each of the three wheels and captures each button press to generate an appropriate system-level event.}
        \label{fig:components}
    \end{minipage}
    %
\end{figure}
\subsection{Hardware Components}
\label{sec:hw}
We designed the 3D components in modules using OpenSCAD~\cite{openscad}. 
The modular parts were printed using a Prusa i3 3D printer with white PLA filament. 
Figure \ref{fig:device3} shows a high-level rendering of the device putting all individual 3D parts together, and Figure \ref{fig:3d-parts} shows a rendering of its eight distinct 3D parts: wheelbase, encoder holder, wheel, connector, main base, button stand, top shell, and button covers. 
Important dimensions are given in millimeters (mm). 
For example, the radius of each wheel is 12 mm, and the width is 10 mm.
We plan to release our design on public repositories for wider adoption.

\subsection{Electrical Components} 
We used an Arduino Pro Micro as the main controller and three rotary encoders to detect rotation in three wheels. 
To provide audio-haptic feedback, we used a buzzer and a pager motor.
The buzzer alerts the user with beep feedback, and the pager motor provides haptic feedback when instructed. 
These electronic components are assembled within the 3D printed parts.
Figure \ref{fig:components} shows a schematic diagram of these components put together---all of which cost less than $\$30$ USD in total.
We will publicly release the model, part number of all electrical components, and the 3D design.

\subsection{\sysname{} Firmware}
Our firmware was developed on the Arduino platform. 
Since Arduino Pro Micro works as a Human Interface Device (HID), \sysname{} does not require an additional device driver to recognize it as a mouse-like device on a computer when connected through a serial port. 
Each rotary encoder has \emph{twelve} slots---each of which provides a clicking effect (mild haptic feedback) in the finger when rotating the wheel. 
In \fnav{} mode, the firmware detects the rotation direction for each wheel, calculates each rotation step, encodes this information into a mouse event, and sends it to the computer, which interprets it as a regular mouse event. 

\new{However, in \hnav{} mode, \sysname{}'s firmware is integrated with NVDA, an open-source screen reader. 
The firmware appears as an NVDA add-on and has access to the UI hierarchy of any app from the NVDA APIs, which internally consume Windows' native UI Automation API~\cite{ui_automation} to extract the UI tree and relay the rotational input on the tree.}

In \tnav{} mode, the firmware issues a mouse event. It consumes an accessibility API call (\texttt{\small AccessibleObjectFromPoint}) to check the closest neighboring UI along the direction of the cursor movement.
%

\section{Evaluation of \sysname{}}
\label{sec:evaluation}
We recruited 12 blind participants and conducted two lab studies to evaluate \sysname{}.
To ensure fair evaluation, we did not invite the four participants who took part in our early design iterations (Sec.~\ref{subsec:design_iterations}).

\textbf{First}, we evaluated the  effectiveness of \wh{}'s \hnav{} mode in \textit{Study 1}, described in Section~\ref{sec:study1}. This study was completed in a single session.
\textbf{Second}, we evaluated \wh{}'s \fnav{} mode in \textit{Study 2}, described in Section~\ref{sec:study2}. 
This study was completed in six sessions and occurred on six different days over a month. 
All studies were IRB-approved.
Table~\ref{table:study} presents an overview of two studies, including tasks, conditions, related hypotheses, and the number of sessions or duration.
Next, we describe \partis{}' demographics and common study procedure, followed by Study 1 and Study 2, and finally, the findings (Sec.~\ref{subsec:findings}). 

\subsection{Participants}
$12$ blind participants ($8$ male, $4$ female) were recruited with an average age of $31.33$  ($SD=5.48$, $Range=22-39$) through an institution that provides services to people with vision impairments.
All of them were familiar with Windows desktop screen readers (e.g., JAWS and NVDA).
None had any motor impairments. 
Some \partis{} had light perception. 
\Partis{} came from a diverse background; most of them were students. 
Table~\ref{table:participants} presents participant demographics and individuals' self-reported expertise with screen readers. 
%

\begin{table*}[t!]
\begin{tabular}{l C{1.0cm}  C{1.8cm} C{1.6cm} C{2.0cm} C{3.2cm}  C{2.30cm}}
 \textbf{ID} & \textbf{Age/ Sex} & \textbf{Expertise} & \textbf{Light Perception?} & \textbf{History of Blindness} & \textbf{Profession} & \textbf{Screen Readers}\\ 
 \toprule
 P1 & 35/M & Expert & No & Advantageous & Student & JAWS, NVDA \\ \hline
 \rowcolor{gray!10} 
 P2 & 25/F & Beginner & No & Congenital & College Student & JAWS, NVDA \\ \hline
 P3 & 32/F &Beginner &Yes &Advantageous & Undergrad student &JAWS, NVDA \\ \hline
 \rowcolor{gray!10} 
 P4 & 38/M &Beginner &No &Advantageous &Undergrad student &JAWS, NVDA \\ \hline
 P5 & 31/F &Beginner &No &Advantageous & Graduate student &JAWS  \\ \hline
 \rowcolor{gray!10} 
 P6 & 31/M &Beginner &No &Advantageous &Education &JAWS  \\ \hline
 P7 & 28/F &Beginner &No &Congential & Special Ed. Trainee &JAWS  \\ \hline 
 \rowcolor{gray!10} 
P8 & 22/F &Beginner & No &Congential & NGO Worker  & JAWS, NVDA  \\ \hline
P9 & 32/M &Beginner & Yes &Congential & Sales &JAWS, NVDA  \\ \hline 
\rowcolor{gray!10} 
P10 & 30/M &Beginner &Yes & Congential & Communication support & NVDA, Magnifier \\ \hline 
P11 & 26/M &Expert &No & Congential & IT program officer & JAWS, NVDA  \\ \hline
\rowcolor{gray!10} 
P12 & 39/M &Expert &No & Advantageous & Small business owner &NVDA, JAWS  \\
 \bottomrule
\end{tabular}
\caption{Participants' demographics, history of blindness, and self-reported expertise with screen readers.} 
\label{table:participants}
\end{table*}
\begin{table*}[t!]

\begin{tabular}{C{2.5cm} C{1cm}  C{1.80cm} C{1.80cm} C{2.0cm} C{1.8cm} }
 \textbf{Study} & \textbf{Task} & \textbf{Conditions} & \textbf{Associated} & \textbf{Design} & \textbf{Sessions/}\\ 
  \textbf{} & \textbf{} & \textbf{} & \textbf{Hypotheses} & \textbf{} & \textbf{Duration} \\
 \toprule
 Study 1: Evaluation of  \hnav{} mode & T1 & C0, C1 & H1 & Within-subject  & 1 Session\\ \hline
 \rowcolor{gray!10} 
 Study 2: Evaluation of  \fnav{} mode & T2 & C2 & H2, H3 & Repeated measures & 6 Sessions\\ \hline

\end{tabular}
\caption{An overview of two studies, including tasks, conditions, related hypotheses, and duration.}
\label{table:study}
\end{table*}

\subsection{Study Procedure}
\label{sec:procedure}
This section describes the procedure that was common in both studies. 
In addition, study-specific procedures are described in the respective study sections. 

The lab studies were conducted in an office environment by two authors. 
After verbal consent, the conductors asked participants to introduce themselves, their history of blindness, their expertise in screen readers (self-disclosed), and their use of point devices and screen reader cursors. 
The experiment was set up on a Windows 10 laptop with $1366\times768$ screen resolution. 
This laptop had the following software installed: two screen readers (JAWS and NVDA), a video conferencing software (Zoom), a remote desktop software (TeamViewer), and the device driver for \sysname{}, which was connected via a USB port.

Of the two authors who administered the study, one was blind and interacted with the participants in person, following social-distancing guidelines. 
The other author was sighted and assisted in setting up the study environment and trials over TeamViewer and supervised each session remotely over Zoom video conferencing software. 
The participants were given sufficient instructions and time ($10 - 30$ minutes) to familiarize themselves with \sysname{}. 
Each session lasted an hour and was video-recorded and later transcribed for further analysis.
Each participant was compensated with an hourly rate of USD \$10.

Upon completing each study, the experimenters engaged in an open-ended discussion, seeking subjective feedback, recommendations, and ratings on different aspects of \sysname{}, such as the placement of wheels, ease of use, the dynamic pace control feature, and perceived challenges in learning this new interaction paradigm and potential of using this device in everyday technology use. 

\subsection{Study 1: Evaluation of \hnav{} Mode}
\label{sec:study1}

In this study, we aim to validate the following hypothesis:
\begin{itemize}
    \item \textbf{H1}: \Partis{} will navigate hierarchical structures more efficiently with \sysname{}'s \hnav{} mode than with a keyboard and a screen reader.
\end{itemize}

\subsubsection{Study Design}
We chose Microsoft Office Suite's ribbon-based multi-level menu as the representative hierarchical structure. 
As reported in prior studies~\cite{haena-speeddial1, haena-speeddial2}, navigating ribbon items is particularly challenging for blind users.
As such, any improvement in navigating ribbons is important to the blind community.

The \partis{} performed the following navigation task (T1) using two study conditions (described below). 
%
\begin{itemize}
    \item[\textbf{Task}] \textit{\textbf{T1}:} In a multi-level hierarchical menu, navigate to a sub-menu item,  given its path in the hierarchy. For example, in Figure~\ref{fig:hnav_illustration}, a representative task could be \textit{``go to \texttt{\small Home} tab, then \texttt{\small Alignment} group, then \texttt{\small Wrap Text} item''}. 
    Here, \texttt{\small Wrap Text} is the target and its path starting from the top, \texttt{\small Home}> \texttt{\small Alignment}> \texttt{\small Wrap Text}, was given to the \partis{}.
\end{itemize}

\noindent The two conditions are described as follows:
\begin{itemize}[leftmargin=5em]

    \item[\textbf{Condition}] \textit{\textbf{C0}: Keyboard with Screen Reader.} The participants must use a screen reader and basic navigational keys, including \texttt{\small ARROW} keys (e.g., $\uparrow$, $\downarrow$, $\leftarrow$, $\rightarrow$) and other modifier keys (e.g., \texttt{\small TAB}, \texttt{\small Alt}, and \texttt{\small ESC}). 
    This was our baseline. 
    \item[\textbf{Condition}] \textit{\textbf{C1}: \sysname{} in \hnav{} mode with TTS.} The participants must use \sysname{}'s \hnav{} mode with a Text-to-Speech (TTS) synthesizer. 
    They were not allowed to use a screen reader or its keyboard shortcuts.
\end{itemize}

\noindent We recorded the task completion times. Five trials were performed in each condition. 

\subsubsection{Procedure Specific to Study 1}
For task T1, we chose ribbon tabs from two commonly used apps in MS Office Suite: MS Word and MS Excel. 
In addition, we selected the ribbon items that blind users use less frequently. 
For instance, our blind author informed us that other blind users are less likely to use  \texttt{\small References}, \texttt{\small Review}, and \texttt{\small Mailings} ribbon tabs in MS Word and \texttt{\small Draw}, \texttt{\small Formula}, \texttt{\small Data} tabs in MS Excel.
We, therefore, included the target from those tabs.
A sample of task T1 was as follows: 
(i) go to \texttt{\small References}> \texttt{\small Footnotes}> \texttt{\small Insert Endnote}; 
(ii) go to \texttt{\small Review}> \texttt{\small Comments}> \texttt{\small Show Comments}; and 
(iii) go to \texttt{\small Formulas}> \texttt{\small Formula Auditing}> \texttt{\small Show Formulas}.

The participants practiced \hnav{} mode on a different hierarchy (e.g., the tree-view of \texttt{Windows Explorer}) to familiarize themselves with the three wheels.
In each trial, the experimenter randomly drew a target (without replacement) from a predefined list of 30 targets. 
Then, the experimenter read out the target and its path and asked the participant to go to that target using a study condition. The experimenter could repeat this information during a trial if asked. 

By default, the \texttt{\small Home} ribbon tab was expanded. Participants were instructed (but not enforced) to start from a ribbon pane they were currently in, as they completed the previous trial. A trial was completed when a \parti{} focused on the target and declared it verbally. 
We counterbalanced the order of conditions across participants. The experimenter took notes during the session. 
The experimenters allocated $3$ minutes for each trial and recorded $120$ data-points (=$12$ participants $\times$ $5$ trials $\times$ $2$ study conditions) for task T1.

\subsection{Study 2: Evaluation of \fnav{} Mode}
\label{sec:study2}
Evaluating \fnav{} mode was less straightforward than \hnav{} because \fnav{} is not directly comparable to a typical mouse.
For example, sighted users can acquire the target visually with a mouse cursor, whereas blind users must be given the target's location (e.g., x, y coordinates) on the screen to acquire with \fnav{} mode.
Similarly, \fnav{} is not comparable to a keyboard either because keyboard-based navigation acquires the target based on its relative position in the abstract UI tree, not its spatial location on the screen. 

Therefore, we created a real-world scenario in which a blind participant and a sighted confederate collaborated remotely on a shared screen. 
The sighted confederate asked the blind peer to move their mouse cursor over a target UI element on the screen. The confederate additionally provided a rough estimation of the target's screen location.
For example, in a remote desktop session shown in Figure~\ref{fig:partially_accessible}.c, the confederate asked the blind peer to move the cursor over \texttt{\small Google Chrome} icon, which is roughly $5\%$ from the left edge and $60\%$ from the top edge.

Based on the above scenario, we defined the following two hypotheses:
\begin{itemize}
    \item \textbf{H2}: Given the spatial coordinate of the target, \partis{} can independently acquire the target with the mouse cursor using \fnav{}.
    \item \textbf{H3}: \fnav{} is easy to learn and use. 
\end{itemize}

\subsubsection{Study Design}
\label{subsec:study_design}
\fnav{} mode is a novel interaction technique to manipulate the mouse cursor. As such, we conducted the study in multiple sessions to measure how well \partis{}'s performance increased in each session. 
More specifically, the study was a repeated-measure design, where each participant performed the following task (T2) using \fnav{} mode in six sessions on six different days over a month. 
%
%

\begin{itemize}
    \item[\textbf{Task}] \textit{\textbf{T2}: Target acquisition}. Move the mouse cursor to a target point, given its coordinates, as shown in Figure~\ref{fig:fnav_illustration} and Figure~\ref{fig:partially_accessible}.c. For example, on Windows Desktop, take the mouse cursor over \texttt{This PC} icon, which is $x\%$ from the left and $y\%$ from the top of the screen.
\end{itemize}
To the best of our knowledge, current screen readers or mouse keys do not allow one to perform the above task. 
Therefore, we had only one condition (C2) for \fnav{} mode and no baseline.
\Partis{} performed T2 in 6 different sessions; thus the session being the independent variable. 
Each session had 6 trials.

\new{Moreover, using percentage coordinates in \fnav{} mode, instead of pixels or inches, makes the location of a UI element agnostic to different screen sizes or resolutions. 
It also helps sighted collaborators estimate and provide the UI location on the screen to their blind counterparts.}

\begin{itemize}[leftmargin=5em]
\setlength{\itemindent}{0em} 
    \item[\textbf{Condition}] \textit{\textbf{C2}: \sysname{} in \fnav{} mode with TTS.} The participants must use \sysname{}'s \fnav{} mode. 
\end{itemize}

Recall that we measured how well \partis{}'s performances improved over sessions (i.e., learning rate) and how well they adopted \fnav{} mode (i.e., user behavior). 
Towards that, we recorded the following measures: (i) task completion time for all trials; (ii) number of times cursor-location is probed; (iii) number of times cursor speed is changed; and (iv) the mean cursor speed.

\subsubsection{Procedure specific to Study 2}
%
In each session, the participants were given sufficient instructions and time ($10$ minutes) to recap \fnav{} mode. 
They practiced on a webpage containing 12 buttons, organized in $3x4$ grids, as shown in Figure~\ref{fig:fnav_illustration}.

For each trial in T2, the sighted experimenter (i.e., the confederate) reoriented  $35$ icons on the study laptop's desktop screen and randomly chose the target icon. 
In addition, the experimenter placed the mouse cursor in the top-left corner of the desktop.
Each icon was a square with dimensions $36px \times 36px$.  
The experimenter read out the name of the target icon and the coordinate of its center in percentages from the left and top edges of the screen and asked the participants to bring the mouse cursor over this target using \wh{}'s \fnav{} mode. 


A trial was completed when a \parti{} brought the cursor over the target icon, and the TTS read out its name.
For each trial, the experimenters allocated 3 minutes.
If a participant failed to complete a trial within the stipulated time limit, it was recorded as incomplete. It was removed from the evaluation \new{because of a lack of valid completion time}.
In sum, out of $432$ data-points (=$12$ participants $\times$ $6$ trials $\times$ $1$ condition  $\times$ $6$ sessions), $336$ were valid for T2. 

\new{The failure cases were primarily due to timeouts, i.e., when a trial exceeded 3 minutes. 
It happened when the target location was non-trivial to estimate or its size was small. 
For example, estimating 77\% from the left is more challenging than estimating 50\% or 80\% from the left. 
In these scenarios, participants often overshot or undershot the target, became more cautious, and decreased the cursor movement speed (\wc{})---all contributing to increased trial time.}



\begin{figure}[t!]
    \centering
    \includegraphics[width=0.5\textwidth]{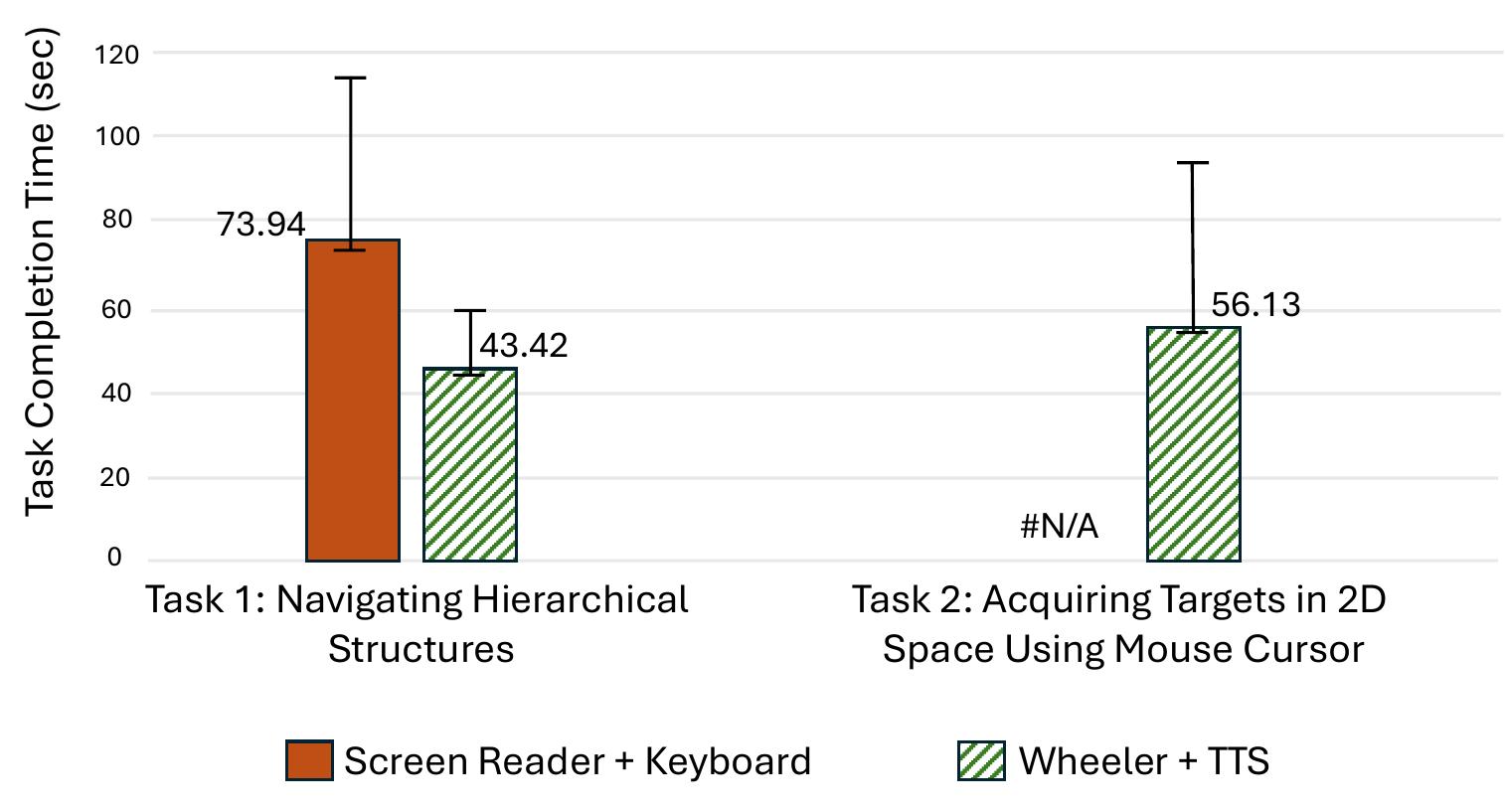}
    \caption{Mean task completion times (the lower, the better) for T1 in two study conditions and T2 in one condition. 
    For T2, no baseline was available, and the reported completion time was observed in the final session. 
    Per session-specific measures are available in Figure~\ref{fig:completion_times_per_session}. 
    Error bars show $+1$ SD.
    }
    \label{fig:completion_times}
\end{figure}

\subsection{Findings from Study-1 and Study-2}
\label{subsec:findings}
%

We analyzed the recorded video data, observations, transcriptions, and experimenters' notes to report our findings. 
Findings from Study 1 appear first, followed by the Study 2.

\subsubsection{Completion Time for Navigating Hierarchical Structure (Study 1)}
\label{subsec:h-nav_eval}
The participants took $40\%$ less time with \sysname{}'s \hnav{} mode ($Mean=43.41s, SD=16.31s$), compared to the baseline ($Mean=73.94s, SD=42.04s$), which is statistically significant, as reported by a paired-t test ($t = 2.303, p < .042$). 
Figure~\ref{fig:completion_times} shows the mean completion time for task T1. 
Recall that the three wheels in \hnav{} mode are mapped to three different hierarchies in the UI tree, and each wheel maintains its state independently of the others. 
For this reason, \partis{} could jump from one sub-tree to another (e.g., between cousin nodes) by simply rotating the first or second wheel. 
In contrast, to make a similar jump with the baseline, they needed to go to the parent node first, then the parent's siblings (uncles), and finally their (uncles') children, which was cumbersome and time-consuming.
Thus, the reduction in completion time in navigating multi-level hierarchies with \hnav{} was expected and unsurprising. 
This validates our hypothesis H1.


\subsubsection{Completion Time in Acquiring Known Targets with the Mouse Cursor (Study 2)}
\label{subsec:2d-nav_eval}
\begin{figure}[t!]
    \centering
    \begin{minipage}{.48\textwidth}
        \centering
        \includegraphics[width=0.94\textwidth]{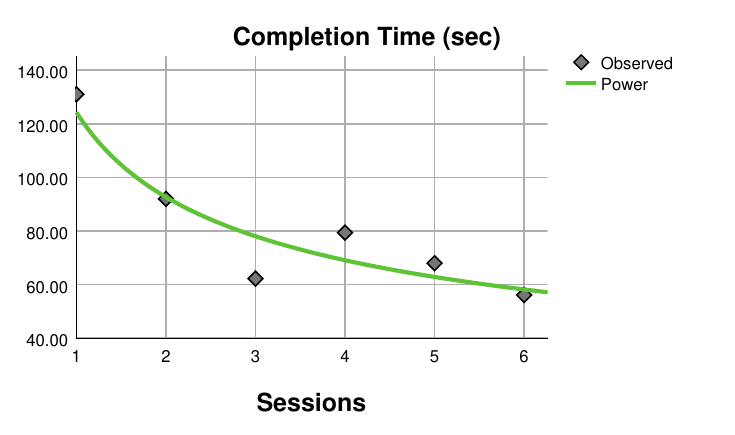}
        \caption{Average task completion time per session for T2 in \fnav{} mode, fitted to a decreasing power trendline, $y=124.28*x^{-0.42}$, $R^2=.83$.}
        \label{fig:completion_times_per_session} 
    \end{minipage}
    \hfill
    \begin{minipage}{.48\textwidth}
        \centering
        \includegraphics[width=0.94\textwidth]{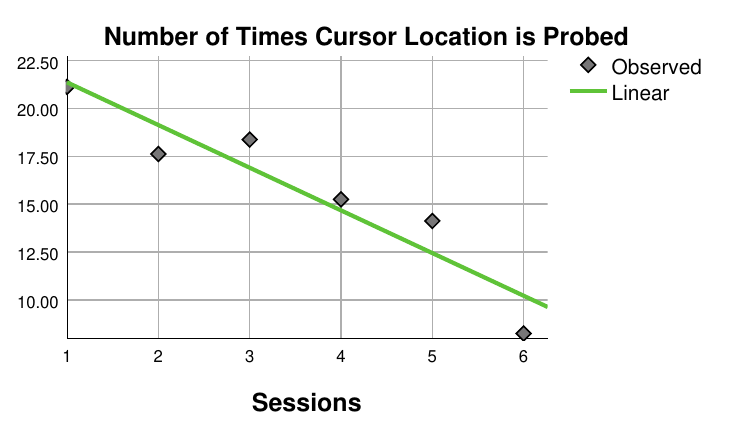}
        \caption{Average number of times cursor location is probed during T2 with \fnav{}, fitted to a linear, decreasing trendline ($R^2 = .883$).}
        \label{fig:cursor_location_probe}
    \end{minipage}
\end{figure}

The task completion time in \fnav{} mode generally follows a decreasing trend over sessions. 
For example, in session 1, the average completion time was $120s$ (shown in Figure~\ref{fig:completion_times_per_session}), which went down to $56.12s$ in session 6 (shown in Figure~\ref{fig:completion_times_per_session} and Figure~\ref{fig:completion_times}). 
These decrements are statistically significant, as indicated by a one-way within-subject ANOVA test, $F(5,60) = 622.4, p \approx 0$.
Although we noticed minor increments in sessions 4 and 5, these could be attributed to a longer interval (e.g., 14 days) between session 3 and session 4, whereas other intervals were 5 to 6 days. 
Overall, the decreasing trend indicates that the task completion time can decrease even further as someone uses \fnav{} consistently. 

\begin{figure}[t!]
    \centering
    \begin{minipage}{.48\textwidth}
        \centering
        \includegraphics[width=0.94\textwidth]{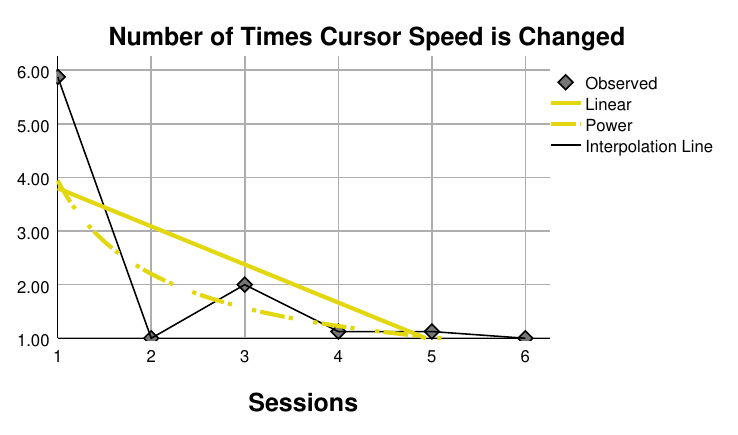}
        \caption{Average number of times cursor speed is changed during T2 with \fnav{}. 
        The plot shows that most \partis{} had settled on a preferable speed after the first 3 sessions.} 
        \label{fig:cursor_speed_change}
    \end{minipage}
    \hfill
    \begin{minipage}{.48\textwidth}
        \centering
        \includegraphics[width=0.94\textwidth]{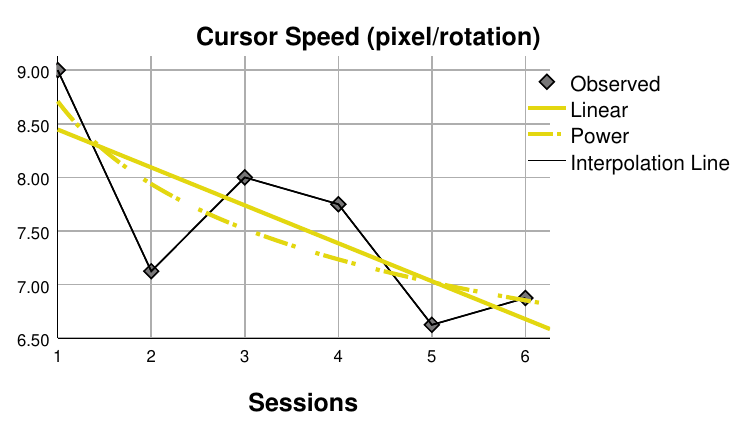}
        \caption{Average cursor speed per session during T2 with \fnav{}. 
        This plot shows that the cursor speed decreased over time, and \partis{} comfortably settled it on 7.0 pixel/rotation speed.}
        \label{fig:pixel_rotation_changes}
    \end{minipage}
\end{figure}

During a task, the \partis{} did not ask for assistance on the whereabouts of their cursor. 
They controlled the cursor pace by using \wc{} (see Figures~\ref{fig:cursor_speed_change} and \ref{fig:pixel_rotation_changes}) and probed the cursor location from time to time by pressing \texttt{\small CTRL} (see Figure~\ref{fig:cursor_location_probe}).
All validate our hypothesis H2.

We noticed that the average completion time of $56.12s$ was still longer than the time sighted users spend acquiring a visual target using a mouse. 
This issue also emerged from the diary study findings with our blind co-author (Section~\ref{sec:diary}). 
Per our blind co-author's recommendation, we created \tnav{}, a special case of \fnav{} mode, to teleport the mouse cursor to the closest neighboring UI along the direction of the cursor movement.

\new{Participants probed the cursor location from time to time (using the \texttt{CTRL} key) to update the corresponding cursor location in their mental map. 
This increased the mental workload for some participants at the beginning. 
However, as they progressed through the study, they increasingly became more comfortable and confident with this mapping, which is evident in Figures~\ref{fig:completion_times_per_session}---\ref{fig:pixel_rotation_changes}.}

%
%
%


\begin{figure}[t!]
    \centering
    \includegraphics[width=0.99\columnwidth]{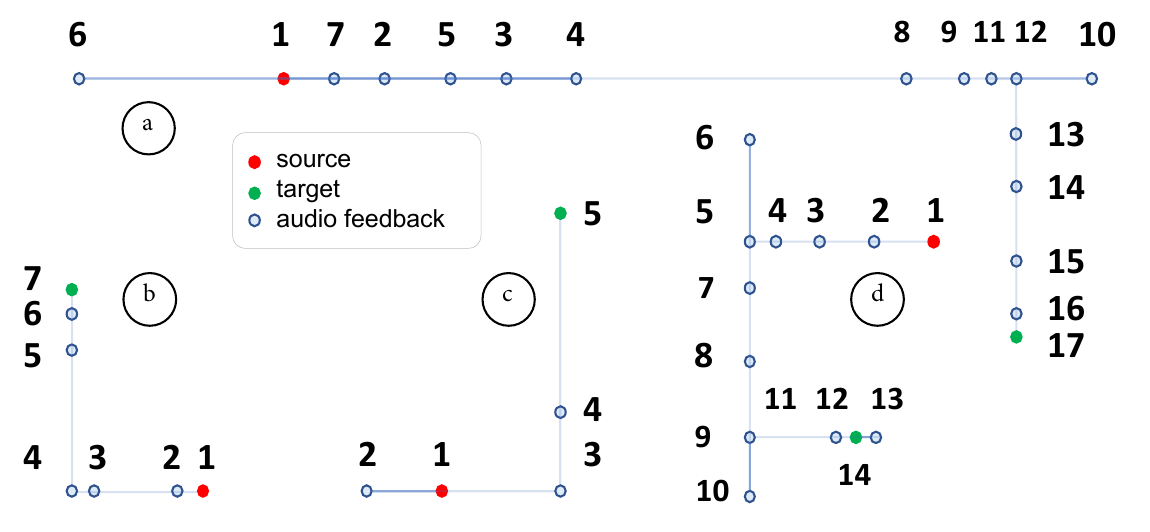}
    \caption{
    Four dominant paths or trajectories that \partis{} followed to move their cursor in \fnav{} mode from a source UI to a destination UI (the target). A red circle (marked with the number 1) is the source, the green circle (marked with the highest number in a path) is the destination, and the blue circles indicate a cursor probe (by pressing \texttt{\small CTRL}) at that location. 
    }
    \label{fig:2d-nav-participants}
\end{figure}

\subsubsection{Learnability in \fnav{} mode}
A key insight gleaned from the data is that in \fnav{} mode, the average task completion time correlated well ($R^2 = 0.83$) with the power law of practice~\cite{SnoddyLearningAS}, as shown in Figure~\ref{fig:completion_times_per_session}), and learning occurred in the last session, suggesting that the participants are more likely to take less time as they practice more.
Another key insight is that the participants oriented themselves with the two-tone audio feedback conveying the cursor's current coordinates, as indicated by the decreasing liner trendline ($R^2 = .883$) in Figure~\ref{fig:cursor_location_probe}.
In addition, they became more comfortable using \fnav{} after multiple sessions and learned how to operate it more efficiently.
For instance, most \partis{} figured out a preferable speed (e.g., 7.0 pixel/rotation) after the first 3 sessions, as shown in Figures~\ref{fig:cursor_speed_change} and Figures~\ref{fig:pixel_rotation_changes}.
These findings suggest that \sysname{} is easy to learn and easy to use. Thus, our hypothesis H3 is validated.



\subsubsection{Movement Trajectories in \fnav{} Mode (Study 2)}
\label{subsec:participant_trajectories}
We recreated the paths or trajectories that \partis{} followed to move their cursor in \fnav{} mode from a source UI to a destination UI (the target). 
Figure \ref{fig:2d-nav-participants} shows four such paths ('a' to 'd'), where a red circle (marked with the number 1) is the source, the green circle (marked with the highest number in a path) is the destination, and the blue circles indicate a cursor probe (by pressing \texttt{\small CTRL}) at that location. 
These trajectories reveal the following insights:

\textit{First}, \partis{} could get confused about moving left or right (or up or down) at the beginning (shown in paths \textit{a} and \textit{c}), but they quickly figured it out. 
For example, in path \textit{a}, the target was on the right side of the source, but a \parti{} initially moved left, realizing the two-tone feedback sounded wrong, then probed the cursor to be certain, and changed the course to the right side (i.e., the correct direction).

\textit{Second}, most \partis{} moved along the X-axis or Y-axis in a long, continuous stride (e.g., path \textit{a}) instead of moving in a staircase pattern (along both axes). This was surprising because we anticipated staircase patterns to be dominant. This also indicates that the \partis{} had developed some notion of spatial awareness of their cursor based on various audio-haptic feedback provided by \sysname{}.

\textit{Third}, as \partis{} approached the target or a turn, they probed the cursor more to be ascertained. Although this behavior was unsurprising, we were surprised by the caution \partis{} took not to overshoot the target.

\noindent These are encouraging insights that indicate the potential of \fnav{} for blind users.

It is evident from the above discussion that locating targets in 2D space using \sysname{}'s \fnav{} mode can yield trajectories different from what a sighted user would take when locating targets using a mouse.
Although the movement time/target locating time can be higher to start with, 
we show in Appendix~\ref{appendix:speed_up} how blind users can achieve sighted user-like performance by changing the cursor pace using \emph{wheel-3} in \fnav{} mode.
The idea presented in Appendix~\ref{appendix:speed_up} is mostly theoretical but can offer intriguing insights into possible improvements blind users can achieve through the long-term use of \sysname{}.

\begin{figure}[ht!]
    \centering
    \includegraphics[width=0.99\columnwidth]{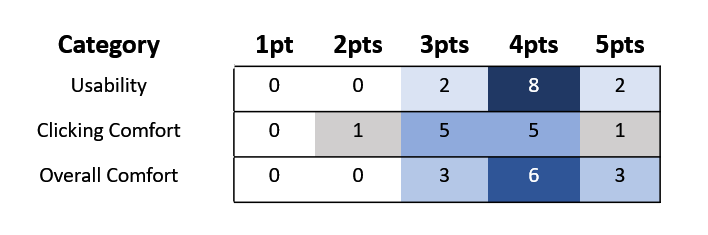}
    \caption{\Partis{} ratings (1 to 5) on the use of comfort of \sysname{} in three categories: usability, clicking comfort of the buttons, and overall comfort or satisfaction. 
    One (1) is very negative, and five (5) is very positive. 
    The shade of a cell indicates the frequency of responses, which is also shown numerically within a cell.
    Note that 4 (i.e., \textit{positive}) is the most frequent response. 
    }
    \label{fig:comfort-ratings}
\end{figure}
\subsection{Observations and Subjective Feedback}
\label{sec:feedback}

\subsubsection{Feedback on Comfort}
\Partis{} rated the use of comfort on a Likert scale of 1 to 5 (1: very negative, 5: very positive) under three categories: 
usability, clicking comfort of the buttons, and the overall comfort or satisfaction of using the device. 
Most \partis{} rated 4, i.e., \textit{positive}, on all three categories, as shown in Figure \ref{fig:comfort-ratings}. 
The mean scores for usability, clicking comfort, and overall comfort were 4.0, 3.5, and 4.0. 
We noted that some \partis{} had difficulty clicking the secondary (small) button. When we asked, they mentioned that the two buttons were placed too closely on the same side. 
Their suggestions and feedback on improving the design are presented in Section~\ref{sec:feedback} below.

Eight participants reported that they found the device very useful. Notably, they liked the idea of moving the cursor with three wheels. Four of them mentioned that they faced difficulties while using it for the first time. However, they remained confident that their experience of using the device would improve over time.

\subsubsection{Feedback on \wh{} Design}
Six participants suggested making the wheels and buttons smoother to grip the device firmly and perform operations more comfortably. 
Four participants mentioned that they found it challenging to use the two buttons under the thumb. 
They suggested increasing the space between the two buttons.  
On the contrary, two participants suggested that placing the secondary button on the other side of the device would be more convenient. 
One participant wished to switch the modes (between \hnav{} and \fnav{}) from the context menu of the secondary button. 
Another participant (P9), who had light perception, suggested that a bright-colored cursor could benefit people with low vision.
Two \partis{} (P6, P8) suggested placing one wheel that moves the cursor along the \texttt{X-axis} horizontally in order to make it more relatable.
However, we argue that placing a wheel horizontally will make it difficult to rotate, thus raising a usability concern.

\subsubsection{\hnav{} vs. \fnav{} mode}
All of our participants mentioned that both modes are useful, complementary, and have distinct use cases. 
For example, 3 \partis{} who frequently use different software in MS Office Suite (e.g., Word, PowerPoint, and Excel)  for employment were enthusiastic about \hnav{} mode. They mentioned that they would never use a keyboard to navigate ribbon menus in MS Office Suite. 
Two other participants mentioned that changing mouse pointer speed, probing the cursor, and two-tone audio feedback helped them picture the spatial layout of the desktop.
They were surprised to discover that the 'Window Start' menu was located in the bottom-left corner for the first time.
P12 provided several use cases where he must need \fnav{} mode: editing an image and working with graphical objects in Unity. 
A very different use case was echoed in P11's comment:

\begin{quotation}
\textit{``... you know many applications and websites have `blindspots' where my keyboard cannot reach, and my screen reader does not talk. I think I can access those blind spots with \sysname{} [\fnav{} mode]''}.
\end{quotation}
\section{Long Term Usability of \sysname{}: A Diary Study}
\label{sec:diary}

\paragraph{Background.} 
After conducting two studies, the blind author (BA) of this paper decided to use \sysname{} on his laptop for everyday technology access. 
BA is a 40-year-old Asian male who runs an institution that provides IT services to blind and low-vision users. 
He is blind by birth and has no light perception.
Also, he is a ``power'' user and comfortable with multiple screen readers (e.g., JAWS, NVDA). 
Although BA does not have a background in computer programming, he understands how screen readers work.

\begin{figure*}
  \centering
  \includegraphics[width=.99\linewidth]{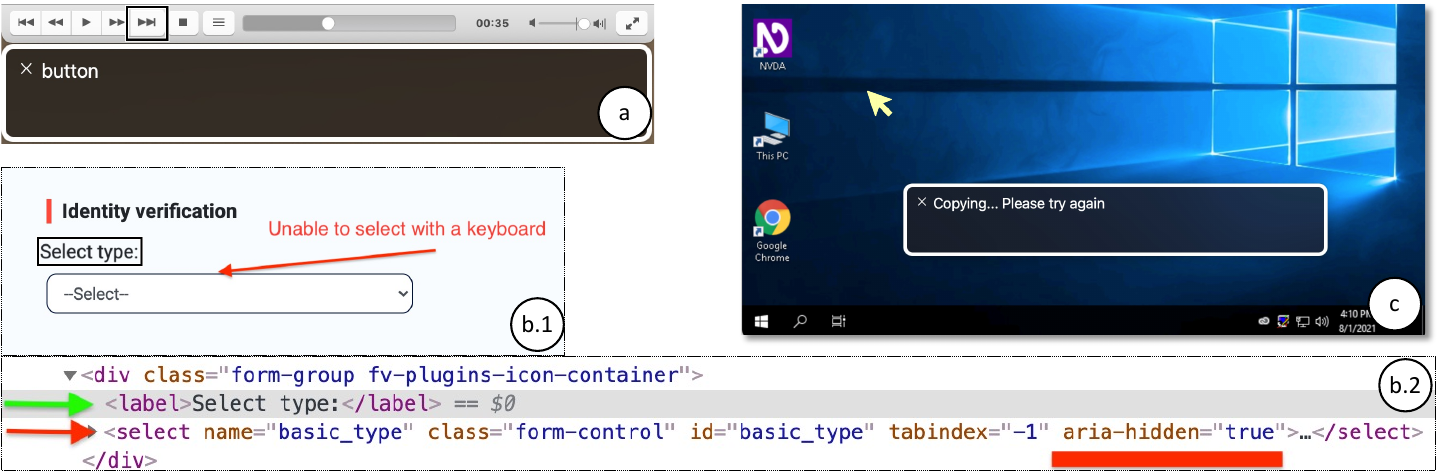}
  \caption[Examples of partially accessible application and website.]{Examples of partially accessible application and website. (a) On the VLC media player in OSX, two image buttons, \texttt{\small Play Next} and \texttt{\small Play Previous}, lack proper labels. As such, the screen reader only reads ``button''. (b.1)  On a COVID-19 vaccine registration page, \texttt{\small Select Type} drop-down list is inaccessible (i.e., hidden) to screen readers because of its incorrect ARIA label, \texttt{\small aria-hidden = ``true''}, as highlighted by a thick red line in its DOM tree shown in (b.2).
  (c) A screenshot of a virtual desktop, which is not accessible to OSX's native screen reader, \texttt{\small VoiceOver}. }
  \label{fig:partially_accessible}
\end{figure*}

BA journaled his experience of using \sysname{} and met other (sighted) authors over Zoom (a teleconferencing software) periodically, e.g., every two weeks for six months.
In each meeting, BA updates his usage pattern and issues (if any) with the device.
Below, we briefly summarize the findings from these meetings. 

\paragraph{\textbf{Usage of \sysname{}}}
BA mentioned that he uses \hnav{} mode extensively, every day, for his job. 
For example, he accesses the hierarchical structures (e.g., ribbon, multi-layer menus) in Word documents, Excel spreadsheets, and Outlook email clients with \hnav{}. 
Therefore, \hnav{} works as envisioned.

BA also mentioned how he used \fnav{} recently to overcome accessibility issues in several scenarios, some of which are shown in Figure~\ref{fig:partially_accessible}: 

\textit{Scenario 1.} When the COVID-19 vaccine was available in his country, he needed to make a reservation by registering online. 
On the registration website, there was a mandatory drop-down (shown in  Figure~\ref{fig:partially_accessible}.b.1) to select his identity type (e.g., passport, birth certificate), which was not reachable to keyboards. 
Still, he managed to access it using \fnav{} mode.
He described that moment as follows: \textit{``No one else was there to help. Everyone was registering for themselves. But luckily, Wheeler worked!''}
When we investigated the website, we found that the drop-down had an incorrect ARIA label, \texttt{\small aria-hidden = ``true''}, making it hidden from the screen reader (shown in  Figure~\ref{fig:partially_accessible}.b.2).

\textit{Scenario 2.} He could click on the playback speed, seek, and pause audio in the VLC Media Player app. These buttons have no text labels and are confusing to navigate with a screen reader (shown in  Figure~\ref{fig:partially_accessible}.a). 

\textit{Scenario 3.} He mentioned that inserting and resizing videos in PowerPoint slides is difficult with keyboard shortcuts, but he found a workaround with \fnav{}.

\textit{Scenario 4.}  He shared an incident on how he troubleshot a client's computer remotely via Zoom screen sharing. His client was unable to access an image button on their computer. 
BA asked the client to grant him remote control permission and then used \fnav{} mode to click on that button for him. In his comment, \textit{``It took some time to figure out where that button was on the screen. But I found it. It felt so good!''}  

\paragraph{\textbf{Feature Request.}}
BA mentioned that he often explores the spatial layout of an app or website using \fnav{}. 
However, he found it  ``too slow'' to move from one element to another, especially when two elements have a large gap.  
So, he suggested an option to jump from one element to another if they are in close proximity.
Based on this request, we implemented \tnav{} mode, which teleports the cursor, as described in Section~\ref{sec:tnav}.
After implementing this feature, BA mentioned that he enjoys using \sysname{} in \tnav{} mode to explore the spatial layout of different apps and teaches his clients and fellow blind users about the basic layout of popular apps/websites.

In sum, our findings from the diary study indicate that \sysname{} can substantially improve the non-visual interaction experience for blind users.

\section{Discussion}
Our findings show that \sysname{} can improve the hierarchical menu navigation in applications with its \hnav{} mode and allows blind users to locate location-known targets in 2D space with its \fnav{} mode.
Here, we discuss the broader implications, limitations, and future directions of \sysname{}.

\subsection{Need for a Pointing Device in Non-Visual Interaction}
For sighted users, interacting with a graphical interface using a mouse is independent of whether individual UI elements contain underlying textual metadata. 
However, this is not the case for blind users because screen readers rely on the metadata to generate audio feedback on the focused element.
Therefore, if an element does not have sufficient textual metadata (e.g., empty \texttt{\small label, alt-text}  attributes~\cite{guinness2018caption, ross2018, Billah_sinter}) or has incorrect attribute value (e.g., \texttt{\small aria-hidden = ``true''} to an otherwise visible web element), the element becomes inaccessible or unreachable to screen readers. 
Unfortunately, these inaccessible UI elements pose a major challenge for blind users~\cite{momotaz_plugin}, for which they seek sighted assistance (e.g., asking sighted family members or calling remote sighted agents~\cite{lee2020emerging} in services like  AIRA~\cite{Aira}, Be My Eyes~\cite{bemyeyes}) or search for a screen reader plug-in~\cite{momotaz_plugin}.
As indicated in the diary study (Section~\ref{sec:diary}), the blind author (BA) of this paper often encountered inaccessible UI elements; 
he successfully interacted with those elements using \sysname{} with limited or no sighted assistance. 
This highlights the need for a pointing device in non-visual interaction and \sysname{}'s potential to fulfill that need. 

\subsection{\textbf{An Augmentation to Keyboard and Screen Reader-based Interaction}}
While \sysname{} provides advantages over traditional non-visual interaction methods like keyboard and screen readers, its purpose is to augment the existing methods, not to replace them.
A notable advantage of screen reader-based interaction would appear when the user knows most shortcuts for navigating an application's items.
Blind users often rely on application shortcuts, but studies indicate limited shortcut knowledge even among experienced blind users~\cite{touhidul2023probabilistic}. 
When shortcuts are unknown, unavailable, or hard to remember, \sysname{} still facilitates faster hierarchy navigation. 
Thus, \sysname{} complements existing methods without aiming to replace them, offering a proficient alternative for efficient non-visual interaction.

\subsection{Increasing Productivity of Blind Users}
For blind users, the inefficiency in operating commonly used office software is a major hindrance to employment~\cite{clements2011factors}. 
To attain basic proficiency with productivity software (e.g., Microsoft Office Suite), blind users typically go through social services and federal- or state-funded specialized training  programs~\cite{employment_outcome_vip}. 
Their training process can be summarized as memorizing numerous keyboard shortcuts and practicing to build muscle memory~\cite{Billah_ubiquitous}.
\sysname{}, as indicated in our findings, can lessen this burden by making access to multi-layer menus fast and structured.

\subsection{Enabling Technology}
Our data suggest that \sysname{} can enable mixed-ability, blind-sighted remote collaboration. For example, a blind user can acquire the target on a shared screen like a sighted user if the target's location is roughly estimated. 
This is important for blind users because the increased acceptance of remote work and improved connectivity software may open new employment opportunities for them~\cite{schur2020telework}.
However, the lack of accessibility to remote collaboration tools is a known issue~\cite{Billah_ubiquitous, telework2021tang, Billah_sinter, das2019doesn}, which can hinder those opportunities~\cite{mack2021mixed}.
\sysname{} can offer an alternative to circumvent these accessibility issues with remote collaboration tools.

Our findings also suggest that \sysname{} can enable working with graphical data, such as editing an image, interacting with online data visualization tools (e.g., Plotly~\cite{plotly}), and 3D modeling with CAD software.  
Although employment opportunities in these areas are increasing, these are largely inaccessible to blind users~\cite{sharif2021understanding, schaadhardt2021understanding, shi2017markit}. Nevertheless, \sysname{}-like devices make them workable for blind users.

\subsection{Potential Application in Virtual Reality}
\sysname{} can be repurposed to be used as an input device in virtual reality (VR).
The current input mechanisms in VR are mostly limited to handheld controllers and gestures---although gesture-based control is still in its infancy for platforms such as Oculus.
Handheld controllers, on the other hand, cannot provide any 3D orientation feedback but provide basic haptics support for events such as boundary hits.
Additionally, there is limited keyboard access and no notion of screen readers in any VR platform, making VR an exclusive domain to sighted users.
With wheeler's \fnav{} mode, accompanied by its feedback mechanism, navigation in the 3D world could be possible for blind users--- 
especially in menu navigation, which is mostly 2D yet lacks any keyboard or screen reader support, something blind users get on desktop platforms such as Windows.

\subsection{Potential Application in Data Visualization}
\new{
Our study findings indicate that \hnav{} mode relies on blind users' ability to conceptualize and navigate hierarchical structures without visual cues. 
This mental mapping enables them to systematically anticipate and traverse different information levels, similar to how sighted users might visually scan and interpret hierarchical data representations.
We believe any data visualization that can be represented as a graph or tree-like form can be traversed conveniently using \sysname{}'s three wheels.
For instance, if the visualization is encoded as a graph, one way to create a hierarchy is to consider the current node as the root, and nodes one hop away from the root are at level 1, two hops away are at level 2, and so on.
Thus, \sysname{}'s 3-wheel architecture provides a versatile framework for navigating various hierarchical data visualizations such as dendrograms, social networks, and organizational charts.}

\new{In dendrograms used for hierarchical clustering, \wa{} would allow users to select the root cluster, while \wb{} would facilitate exploration of immediate clusters and \wc{} wouls help dive deeper into sub-clusters.
For social networks, \sysname{} would enable users to start from a central node (e.g., an influencer or organization) using \wa{}. 
\wb{} then would allow exploration of immediate connections (e.g., friends or followers), while \wc{} would extend exploration to secondary connections, providing insights into community structures and information flow across the network.
In organizational charts, users can begin with the CEO or top-level executive using \wa{}. 
\wb{} then would enable exploration of direct reports and major departments, while \wc{} would allow deeper dives into teams and divisions within each department. 
This structured navigation aids in understanding reporting relationships, departmental structures, and organizational hierarchies.}

\subsection{Limitations and Future Work}
The findings from our diary study (Sec.~\ref{sec:diary}) indicate \sysname{}'s room for improvement, particularly regarding hardware design, such as the placement of the two side buttons.
In future work, we plan to experiment with different placements and sizes of the primary and secondary buttons.
Another limitation of \sysname{}'s \fnav{} mode is the assumption that users have prior knowledge of their desired target location within the UI, either from experience or from a sighted confederate.
One potential solution is integrating an AI assistant capable of receiving spoken instructions from the user and providing tentative screen coordinates for the desired target.

In the future, we plan to write a separate device driver for \sysname{} to broadcast rotational events system-wide so that applications can consume these events directly, similar to standard mouse/keyboard events, thus augmenting the input space for non-visual interaction. 
This can also eliminate writing the application-specific adaptation for rotational inputs.
Moreover, we plan to integrate \sysname{} with an open-source screen reader, such as NVDA~\cite{nvda-github}, to make the transition between keyboard-based interaction to \sysname{}-based interaction seamless. 
Finally, we will make \sysname{} prototype open source by releasing the 3D design, schematic diagram, and part numbers of various electrical components for wide adoption.

\section{Conclusion}
This paper presents a three-wheel mouse-shaped stationary input device, \sysname{}, to make non-visual interaction efficient and versatile. 
This device adopts a rotational input paradigm that prior work has found helpful for blind users.
Informed by prior work, the design of \sysname{} is refined by participatory design sessions and the experience of a blind co-author in this paper. 
The prototype is made of 3D-printed components and commercially available electrical components. 
\sysname{} is evaluated by 12 blind participants in two user studies.
The study findings suggest that \sysname{} can take up to $40\%$ less time navigating dense, hierarchical UI structures.
Moreover, blind \partis{} can maneuver the mouse cursor to acquire a target on the screen given its location.
Further, \sysname{} is easy to use and learn, and users' performance can improve over time.
A diary study with our blind co-author indicates that the device works as envisioned by large.
It can increase the productivity of blind users in using office software and offer several serendipitous benefits, including remote collaboration, interacting with partially inaccessible applications and websites, and promoting independence.

\bibliographystyle{ACM-Reference-Format}
\bibliography{Bibliography, Bibliography2, Bibliography3}

\appendix
\section{\sysname{}'s Design Iterations}
\label{appendix:design_iterations}

As mentioned in Section~\ref{sec:overview}, the version of \sysname{} we presented in this paper has been a result of three major design iterations.
In each iteration, we mirrored the workshop's setup. We briefed participants on the device's objective, notes from previous meetings, and the device’s current status. 
We handed them the in-progress prototype and asked them to navigate the Windows Directory Tree on a computer, first with a screen reader, and then with the device.

Toward the end of the session, we discussed any issues observed and queried the reasoning behind their actions. 
We also asked for general feedback about the device's current state. 
We then prioritized the next steps based on the gathered information, making adjustments to the design and adding/removing electronic components.

In the first iteration, we incorporated 3 wheels, 2 buttons, and a buzzer into the base device. 
In the second iteration, we added haptic feedback and replaced the Arduino Uno with the Arduino Micro, as the former could not issue a mouse hardware event. 
In the 3rd iteration, we included the \fnav{} mode, a toggle event to switch between \hnav{} and \fnav{} modes, text-to-speech readout of the mouse cursor's (x,y) location upon pressing the \texttt{CTRL} button on the keyboard, and a two-tone audible to indicate the mouse cursor’s current position relative to width and height of the screen.

\section{Movement Time Mathematical Formulations}
\label{appendix:MT}

\subsection{Acquiring Targets Using the Shortest Distance}

\begin{figure*}[!ht]
\begin{minipage}[t]{0.4\textwidth}
  \centering
  \includegraphics[width=\columnwidth]{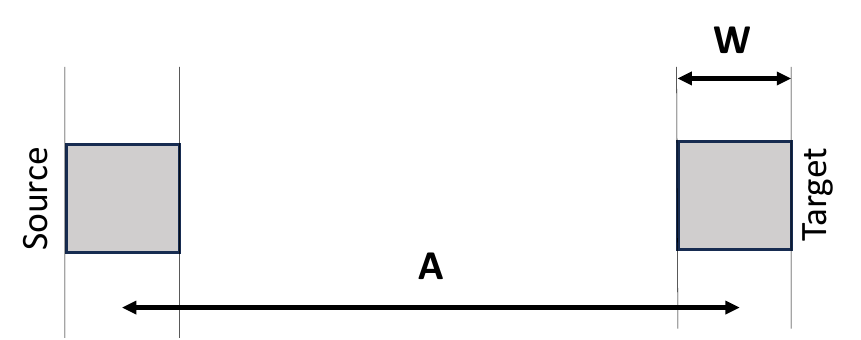}
  \caption{Acquiring Targets Using Shortest Distance.}
  \label{fig:basic_fitts}
\end{minipage}
\hfill
\begin{minipage}[t]{0.4\textwidth}
  \centering
  \includegraphics[width=\columnwidth]{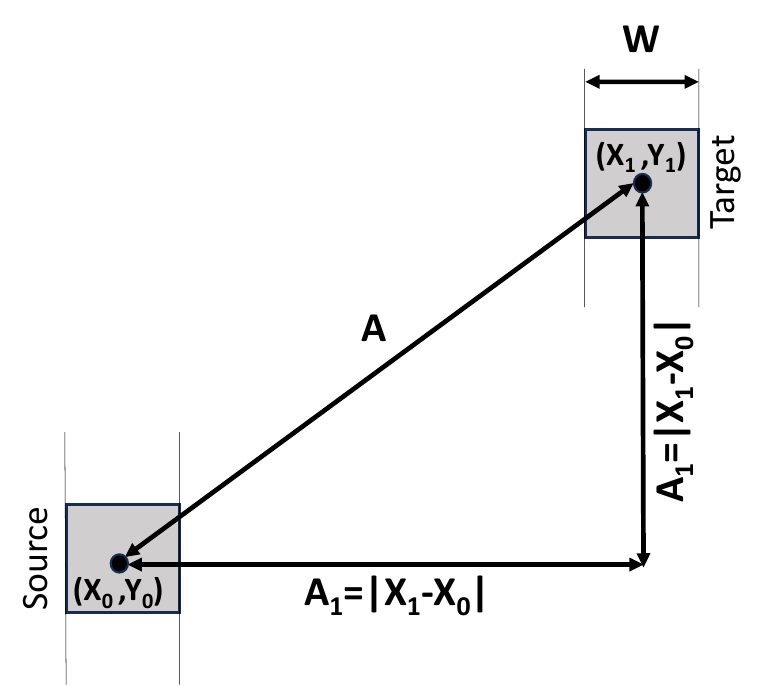}
  \caption{Acquiring Targets Using Rectilinear Movements.}
  \label{fig:rectilinear}
\end{minipage}  
\end{figure*}

Let us consider the scenario presented in Figure~\ref{fig:basic_fitts}; where the user is trying to move the cursor from the source to the target object.
Here, A is the distance between the source and the target, and W is the width of the target.
Jagacinski et al. \cite{jagacinski2018control} presented a derivation of the required time for such a move using the first-order lag system of Control Theory. Here is what the formulation looks like:

\begin{equation}
\label{eq:mt_from_control_theory}
    t = \frac{\ln{(2)}}{k}\log_2{(\frac{2A}{W})}
\end{equation}

Where, k is called the gain factor, which determines the speed at which the target is acquired~\cite{jagacinski2018control}.

Equation~\ref{eq:mt_from_control_theory} is indeed another formulation of the Fitts' law~\cite{fitts1954information}. According to Fitts' law:

\begin{equation}
\label{eq:fitts_law_1}
    t = a + b * ID
\end{equation}

Where ID stands for Index of Difficulty and,  
\begin{equation}
\label{eq:fitts_law_2}
    ID = \log_2{(\frac{2A}{W})}
\end{equation}

\subsection{Acquiring Targets Using Rectilinear Distances}

In \sysname{}'s \fnav{} mode, users use rectilinear movements to go from source ($X_0$, $Y_0$) to target ($X_1$, $Y_1$). 
For example, In Figure~\ref{fig:rectilinear}, the users would travel the distances $A_1$ and then $A_2$, instead of only $A$ like sighted users usually do.
In this section, we try to formulate an equation for movement time using such movements. 

In this scenario, we can derive the movement times for $A_1$ and $A_2$ separately using Equation~\ref{eq:mt_from_control_theory}.
If $t_1$ and $t_2$ unit times are needed for traversing distances $A_1$ and $A_2$ respectively, we can write:

\begin{equation}
    t_1 = \frac{\ln{(2)}}{k}\log_2{(\frac{2A_1}{W})}    
\end{equation}

and, 

\begin{equation}
    t_2 = \frac{\ln{(2)}}{k}\log_2{(\frac{2A_2}{W})}    
\end{equation}

If total movement time in this case is $T_{rec}$, then, 

\begin{equation}
\label{eq:rectilinear_travel}
\begin{aligned}
        T_{rec} &= t_1 + t_2 \\
          &= \frac{\ln{(2)}}{k} [ \log_2{(\frac{2A_1}{W})} + 
          \log_2{(\frac{2A_2}{W})} ] \\
          &= \frac{\ln{(2)}}{k} [\log_2{(\frac{4A_1A_2}{W^2})}]
\end{aligned}
\end{equation}

Where, $A_1=|X_1-X_0|$ and $A_2=|Y_1-Y_0|$.

\subsection{Rectilinear Vs. Shortest Path Travel}

In Figure~\ref{fig:rectilinear}, If path $A$ was taken instead of \{$A_1$, $A_2$\} to reach the target, then the movement time $T_{shortest}$ would be:

\begin{equation}
\label{eq:shortest_path_travel}
    T_{shortest} = \frac{\ln{(2)}}{k}\log_2{(\frac{2A}{W})}
\end{equation}

From Figure~\ref{fig:rectilinear}, it is easy to see that $T_{shortest}$ is smaller than $T_{rec}$.
The following calculation shows the difference between $T_{rec}$ and $T_{shortest}$:

\begin{equation}
\label{eq:delta}
\begin{aligned}
        \Delta T  &= T_{rec} - T_{shortest} \\
        & = \frac{\ln{(2)}}{k} [\log_2{(\frac{4A_1*A_2}{W^2})}] 
        - \frac{\ln{(2)}}{k}\log_2{(\frac{2A}{W})}
        \\
        &= \frac{\ln{(2)}}{k} [\log_2{(\frac{4A_1*A_2}{W^2})} - \log_2{(\frac{2A}{W})}] \\
        &= \frac{\ln{(2)}}{k} [\log_2{(\frac{4A_1*A_2}{W^2}*\frac{W}{2A}})] \\
        &= \frac{\ln{(2)}}{k}[\log_2{(\frac{2A_1*A_2}{AW})}] \\
\end{aligned} 
\end{equation}

Using the Pythagorean theorem, $A$ can be written in terms of $A_1$ and $A_2$ as follows: $\sqrt{A_1^2 + A_2^2}$. Therefore, Eq.~\ref{eq:delta} can be written as below:

\begin{equation}
\label{eq:delta2}
\begin{aligned}
        \Delta T  &= T_{rec} - T_{shortest} \\
        &= \frac{\ln{(2)}}{k}[\log_2{(\frac{A_1*A_2}{A}* \frac{2}{W})}] \\
        &= \frac{\ln{(2)}}{k}[\log_2{(\frac{A_1*A_2}{\sqrt{A_1^2 + A_2^2}}* \frac{2}{W})}] \\
        &= \frac{\ln{(2)}}{k}[\log_2{(\frac{1}{\sqrt{\frac{1}{A_1^2} + \frac{1}{A_2^2}}}* \frac{2}{W})}] \\
\end{aligned} 
\end{equation}

One can interpret the above Eq.~\ref{eq:delta2} as follows: 
\begin{itemize}
    \item If $A_1 \to 0$ or $A_2 \to 0$, i.e., both the source and the target are parallel to X- or Y-axis, the difference in time for rectilinear and shortest movements reduces to zero ($\Delta T \to 0$).
    \item If $A_1 \to A_2$ or $A_2 \to A_1$, indicating the angle between X- or Y-axis and the line connecting the source and the target becomes 45 degrees, $\Delta T$ reaches its maximum value, i.e., $\frac{\ln{(2)}}{k}[\log_2{(\frac{A_1}{\sqrt{2}}* \frac{2}{W})}]$ or $\frac{\ln{(2)}}{k}[\log_2{(\frac{A_2}{\sqrt{2}}* \frac{2}{W})}]$.
    \item If one component is significantly larger than other, $A_1 >> A_2$, $\Delta T$ depends on the smaller component \\($\Delta T \approx \frac{\ln{(2)}}{k}[\log_2{(A_2 * \frac{2}{W})}]$, assuming $\frac{1}{A_1^2} \to 0$).
\end{itemize}

\subsection{Rectilinear Travel at different Speeds}
\label{appendix:speed_up}

\sysname{} has the option to modify the cursor speed (via the third wheel) in \fnav{} mode. This would enable the users to speed up the movements in rectilinear motions if they want to.
If the user speeds up the cursor movement by a factor of $s$ (\textbf{assuming s>1}), then the effective distances they have to travel become $\frac{A_1}{s}$ and $\frac{A_2}{s}$ instead of $A_1$ and $A_2$ respectively. 
If we call the movement time $T_{rec-speed}$ this time, Equation~\ref{eq:rectilinear_travel} becomes:

\begin{equation}
\label{eq:rectilinear_travel_speed}
\begin{aligned}
        T_{rec-speed}
          &= \frac{\ln{(2)}}{k} [ \log_2{(\frac{\frac{2A_1}{s}}{W})} + 
          \log_2{(\frac{\frac{2A_2}{s}}{W})} ] \\
          &= \frac{\ln{(2)}}{k} [ \log_2{(\frac{2A_1}{Ws})} + 
          \log_2{(\frac{2A_2}{Ws})} ] \\
          &= \frac{\ln{(2)}}{k} [\log_2{(\frac{4A_1A_2}{W^2s^2})}]
\end{aligned} 
\end{equation}

Now, let us consider that we pick $s$ in a way so that $T_{rec-speed}$ and $T_{shortest}$ becomes equal. In other words, we want to see how much we have to speed up (s) the movement to achieve the shortest path (sighted) performance from the rectilinear path.

Hence,

\begin{equation}
\label{eq:deriving s}
\begin{aligned}
        & T_{rec-speed}
          = T_{shortest} \\
        &\implies \frac{\ln{(2)}}{k} [\log_2{(\frac{4A_1A_2}{W^2s^2})} =
        \frac{\ln{(2)}}{k}\log_2{(\frac{2A}{W})} \\
        &\implies \frac{4A_1A_2}{W^2s^2} = \frac{2A}{W} \\
        &\implies As^2W = 2A_1A_2 \\
        &\implies s = \sqrt{\frac{2A_1A_2}{AW}} \\
\end{aligned} 
\end{equation}


Where, \\
$A_1=|X_1-X_0|$,  $A_2=|Y_1-Y_0|$, and $A=\sqrt{(X_1-X_0)^2 + (Y_1-Y_0)^2}$

Thus, if we have :

\begin{enumerate}
    \item the source ($X_0$, $Y_0$) and the target ($X_1$, $Y_1$) coordinates, and
    \item  the width of the target (W),
\end{enumerate}

we can determine how much speed increment in rectilinear movements is necessary to achieve the same performance as that of taking the shortest path.

\paragraph{\textbf{When s<1:}}
s=1 means we expect the speed in rectilinear movements to be the same as the shortest path movement.
However, with some users, the conditions s=1 or s>1 may not be achievable.
In \sysname{}'s \fnav{} mode, the third wheel also allows the users to slow down the cursor i.e., activate a scenario where s<1. 
When they do so, the movement time will be higher as indicated by Equation~\ref{eq:rectilinear_travel_speed}.  
\end{document}